\documentclass[12pt]{article}
\usepackage{amsmath,amssymb,amsthm,amsxtra,overpic,bbm,bm,epsfig,subfigure}
\usepackage{hyperref}
\usepackage{mathrsfs}
\usepackage{enumitem}
\usepackage{graphicx}
\usepackage{color}
\usepackage[table]{xcolor}
\usepackage{comment}
\usepackage{epstopdf}
\usepackage{float}
\usepackage{cite}
\usepackage{slashed,stmaryrd}

\allowdisplaybreaks[1]

\def\thefootnote{\fnsymbol{footnote}}

\usepackage{multirow}
\newcommand{\Dlr}{\mbox{$\raisebox{2mm}{\boldmath ${}^\leftrightarrow$}\hspace{-4mm}D^{}_\mu$}}
\newcommand{\Dilr}{\mbox{$\raisebox{2mm}{\boldmath ${}^\leftrightarrow$}\hspace{-4mm}D^I_\mu$}}

\newcommand{\Dl}{\mbox{$\raisebox{2mm}{\boldmath ${}^\leftarrow$}\hspace{-4mm}D^{}_\mu$}}

\newcommand{\Dlrn}{\mbox{$\raisebox{2mm}{\boldmath ${}^\leftrightarrow$}\hspace{-4mm}D^\nu$}}
\newcommand{\Dilrn}{\mbox{$\raisebox{2mm}{\boldmath ${}^\leftrightarrow$}\hspace{-4mm}D^{I\nu}$}}

\newcommand{\Yl}{Y^{}_l}

\newcommand{\Yu}{Y^{}_{\rm u}}

\newcommand{\Yd}{Y^{}_{\rm d}}

\newcommand{\Yli}[2][\alpha]{\left(Y^{}_l\right)^{}_{#1 #2}}

\newcommand{\Yui}[2][\alpha]{\left(Y^{}_{\rm u}\right)^{}_{#1 #2}}

\newcommand{\Ydi}[2][\alpha]{\left(Y^{}_{\rm d}\right)^{}_{#1 #2}}

\newcommand{\Lelli}[1][\beta]{\ell^{}_{#1 \rm L}}
\newcommand{\BLelli}[1][\alpha]{\overline{\ell^{}_{#1 \rm L}}}
\newcommand{\REi}[1][\beta]{E^{}_{#1 \rm R}}

\newcommand{\BLQi}[1][\alpha]{\overline{Q^{}_{#1 \rm L}}}
\newcommand{\RUi}[1][\beta]{U^{}_{#1 \rm R}}

\newcommand{\RDi}[1][\beta]{D^{}_{#1 \rm R}}

\newcommand{\rmI}{{\rm i}}

\newcommand{\Op}{\mathcal{O}}

\usepackage{ulem}

\begin{document}

\begin{center}
{\Large\bf One-loop Matching of the Type-II Seesaw Model onto \\
the Standard Model Effective Field Theory}
\end{center}

\vspace{0.2cm}

\begin{center}
{\bf Xu Li }~\footnote{E-mail: lixu96@ihep.ac.cn},
\quad
{\bf Di Zhang }~\footnote{E-mail: zhangdi@ihep.ac.cn},
\quad
{\bf Shun Zhou }~\footnote{E-mail: zhoush@ihep.ac.cn (corresponding author)}
\\
\vspace{0.2cm}
{\small
Institute of High Energy Physics, Chinese Academy of Sciences, Beijing 100049, China\\
School of Physical Sciences, University of Chinese Academy of Sciences, Beijing 100049, China}
\end{center}

\vspace{1.5cm}

\begin{abstract}
In this paper, we continue to construct the low-energy effective field theories (EFTs) of the canonical seesaw models, which are natural extensions of the Standard Model (SM) to accommodate tiny but nonzero neutrino masses. Different from three right-handed neutrino singlets in the type-I seesaw model, the Higgs triplet in the type-II seesaw model participates directly in the electroweak gauge interactions, rendering the EFT construction more challenging. By integrating out the heavy Higgs triplet in the functional-integral formalism, we carry out a complete one-loop matching of the type-II seesaw model onto the so-called Standard Model Effective Field Theory (SMEFT). It turns out that 41 dimension-six operators (barring flavor structures and Hermitian conjugates) in the Warsaw basis of the SMEFT can be obtained, covering all those 31 dimension-six operators in the case of type-I seesaw model. The Wilson coefficients for 41 dimension-six operators are computed up to $\mathcal{O}\left( M^{-2}_\Delta \right)$ with $M^{}_\Delta$ being the mass scale of the Higgs triplet. Moreover, the branching ratios of rare radiative decays of charged leptons $l^-_\alpha \to l^-_\beta + \gamma$ are calculated in the EFT and compared with that in the full theory in order to demonstrate the practical application and the correctness of our EFT construction.
\end{abstract}

%\begin{flushleft}
%\hspace{0.8cm} PACS number(s):
%\end{flushleft}

\def\thefootnote{\arabic{footnote}}
\setcounter{footnote}{0}
\newpage
%\tableofcontents

\section{Introduction}\label{sec:intro}

The experimental discovery of neutrino oscillations provides us with very compelling evidence that neutrinos are actually massive and lepton flavors are significantly mixed~\cite{Xing:2020ijf}. The origin of neutrino masses and flavor mixing definitely calls for new physics beyond the Standard Model (SM). On the other hand, the SM has so far been the most successful theory for strong, weak and electromagnetic interactions, passing essentially all the experimental tests except for neutrino oscillations~\cite{Zyla:2020zbs}. Such a situation may hint at the widely-accepted idea that the SM just serves as an effective field theory (EFT) at the low-energy scale (i.e., the electroweak scale $\Lambda^{}_{\rm EW} \equiv 10^2~{\rm GeV}$), where the higher-dimensional operators composed only of the SM fields respect the SM gauge symmetry and take the responsibility for all the observed deviations from the SM predictions. 

As first pointed out by Steven Weinberg~\cite{Weinberg:1979sa}, the dimension-five operator in the Standard Model Effective Field Theory (SMEFT) is unique and leads to the generation of tiny Majorana neutrino masses after the spontaneous breakdown of the SM gauge symmetry. More explicitly, the effective Lagrangian of the SMEFT can be written as
\begin{eqnarray}
{\cal L}^{}_{\rm SMEFT} = {\cal L}^{}_{\rm SM} + \sum_i \frac{C^{(5)}_i O^{(5)}_i}{\Lambda} + \sum_i \frac{C^{(6)}_i O^{(6)}_i}{\Lambda^2} + \cdots \; ,
\label{eq:LagSMEFT}
%     (1)
\end{eqnarray}
where $\Lambda$ is the cutoff energy scale for the SMEFT, ${\cal L}^{}_{\rm SM}$ stands for the SM Lagrangian, $O^{(5)}_i$ and $O^{(6)}_i$ are dimension-five (dim-5) and dimension-six (dim-6) operators with $C^{(5)}_i$ and $C^{(6)}_i$ being the associated Wilson coefficients, respectively. The dim-6 operators in the SMEFT have already been systematically studied in Ref.~\cite{Buchmuller:1985jz}, and recently revised in Ref.~\cite{Grzadkowski:2010es}, where the well-known Warsaw basis of 59 independent baryon-number-conserving dim-6 operators (plus four baryon-number-violating operators) has been established. In the precision era of particle physics, the SMEFT obviously offers an extraordinarily useful and efficient way, which is independent of any specific ultraviolet (UV) models, to probe new physics~\cite{Brivio:2017vri}. Recent years have seen tremendous progress in the developments of the SMEFT itself~\cite{Lehman:2014jma, Liao:2016hru, Li:2020gnx, Murphy:2020rsh, Li:2020xlh, Liao:2020jmn, Li:2020tsi,
Jenkins:2013zja, Jenkins:2013wua, Alonso:2013hga, Liao:2019tep, David:2020pzt, Henning:2015alf, Antusch:2001ck, Chala:2021juk} and its many interesting extensions~\cite{Aparici:2009fh, delAguila:2008ir, Bhattacharya:2015vja, Liao:2016qyd, Bischer:2019ttk, Li:2021tsq}. However, as the mass dimension of operators under consideration becomes higher, the number of independent operators in the SMEFT will increase very rapidly~\cite{Henning:2015alf}, rendering a complete experimental determination of all the relevant coefficients associated with the operators to be extremely difficult or even impossible.

For this reason, we take another distinct attitude to the exploration of possible new physics beyond the SM. If the renormalizable UV model is believed to exist, one can choose one of the well-motivated UV models and match it onto the SMEFT by integrating out the heavy degrees of freedom. In this way, only a fraction of the effective operators in the SMEFT will be obtained and the corresponding Wilson coefficients are highly correlated. Then the constructed EFT is confronted with the precision data from various experiments. Apparently the disadvantage of such an approach is lacking of the model independence, which is the primary motivation for the SMEFT. In light of the discovery of neutrino oscillations, it is reasonable to argue that the canonical seesaw models for tiny Majorana neutrino masses are strongly motivated, and the EFTs of these renormalizable UV models will be indispensable for self-consistent phenomenological studies at the low-energy scale.

The basic strategy for the construction of EFTs from renormalizable UV theories has been outlined in Ref.~\cite{Weinberg:1980wa}. By integrating out the heavy degrees of freedom, one can match the UV model onto the SMEFT at the tree level, which has been accomplished in Ref.~\cite{deBlas:2017xtg} for general field content and arbitrary types of interactions. At the one-loop level, a number of examples can be found in the literature~\cite{Bilenky:1993bt, Boggia:2016asg, Ellis:2017jns, Jiang:2018pbd, Haisch:2020ahr, Cohen:2020fcu,Dittmaier:2021fls, Henning:2016lyp, Ellis:2016enq, Fuentes-Martin:2016uol, delAguila:2016zcb, Chala:2020vqp, Gherardi:2020det, Zhang:2021tsq} either by the diagrammatic method or by the functional approach~\cite{Henning:2014wua, Drozd:2015rsp, Ellis:2016enq, Henning:2016lyp, Fuentes-Martin:2016uol, Zhang:2016pja, DasBakshi:2018vni, Ellis:2017jns, Kramer:2019fwz, Cohen:2019btp, Cohen:2020fcu, Cohen:2020qvb, Fuentes-Martin:2020udw,Dittmaier:2021fls,Zhang:2021jdf,Brivio:2021alv,Dedes:2021abc}. In the previous work~\cite{Zhang:2021jdf}, we have initiated a program of performing a complete one-loop matching of the seesaw models onto the SMEFT. For the type-I seesaw model~\cite{Minkowski:1977sc, Yanagida:1979as, GellMann:1980vs, Glashow:1979nm, Mohapatra:1979ia}, the one-loop matching has been achieved and shown to be useful to investigate the radiative decays of charged leptons $l^-_\alpha \to l^-_\beta + \gamma$ in a self-consistent way~\cite{Zhang:2021tsq, Xing:2020ivm}. In the present paper, we continue to carry out the complete one-loop matching for the effective operators up to dim-6 by integrating out the heavy Higgs triplet in the type-II seesaw model~\cite{Konetschny:1977bn, Magg:1980ut, Schechter:1980gr, Cheng:1980qt, Mohapatra:1980yp, Lazarides:1980nt} with the functional approach~\cite{Gaillard:1985uh, Chan:1986jq, Cheyette:1987qz}. The main motivation for such a study is two-fold. First, the Higgs triplet in the type-II seesaw model transforms non-trivially under the SM gauge group, implying a more challenging construction of the EFT than that in the type-I seesaw model. Second, it is intriguing to see how many and what kind of dim-6 operators in the Warsaw basis of the SMEFT can be obtained in the type-II seesaw EFT (SEFT-II). The results should be compared to those in the type-I seesaw EFT (SEFT-I), as derived in Ref.~\cite{Zhang:2021jdf}. The dim-6 operators that are not commonly shared by the SEFT-I and the SEFT-II may indicate which kind of signatures in the precision measurements at the low-energy scale could serve as a discriminator for the type-I and type-II seesaw models as the renormalizable UV theory for neutrino masses. Similar studies can also be extended to the type-III seesaw model~\cite{Foot:1988aq} and other neutrino mass models~\cite{Coy:2021hyr}.

The remaining part of this paper is structured as follows. In Sec.~\ref{sec:framework}, we briefly recall the functional approach to the tree-level and one-loop matching of an UV model onto the SMEFT. In Sec.~\ref{sec:model}, we introduce the type-II seesaw model in order to establish our notations and conventions, and further derive the operators up to dim-6 and the associated Wilson coefficients in the SEFT-II at the tree level. Sec.~\ref{sec:oneloop} is devoted to the one-loop matching of the type-II seesaw model onto the SMEFT by evaluating the supertraces via the publicly available package {\sf SuperTracer}~\cite{Fuentes-Martin:2020udw}. The dim-6 operators and the corresponding Wilson coefficients in the Warsaw basis are given. The results are also cross-checked by the diagrammatic approach. In Sec.~\ref{sec:muegamma}, as an illustrative example for the consistency between the UV model and the EFT, the branching ratios for the radiative decays of charged leptons $l^-_\alpha \to l^-_\beta + \gamma$ are computed in the SEFT-II and compared to the results in the heavy-triplet limit of the full type-II seesaw model. Finally, we summarize our main conclusions in Sec.~\ref{sec:summary}.

\section{The Functional Approach}\label{sec:framework}

The functional approach will be implemented to perform the tree-level and one-loop matching in the present work. In this section, following Ref.~\cite{Zhang:2021jdf}, we briefly explain the general procedure to achieve this goal, and a more detailed account can be found in the literature, e.g., Refs.~\cite{Cohen:2020fcu, Zhang:2016pja}. The basic idea to match a given UV theory onto the low-energy EFT is to identify the one-light-particle-irreducible (1LPI) effective action (i.e., $\Gamma^{}_{\rm L, UV}$) in the UV theory with the one-particle-irreducible (1PI) effective action (i.e., $\Gamma^{}_{\rm EFT}$) in the low-energy EFT at the matching scale, namely,
\begin{eqnarray}
	\Gamma^{}_{\rm L, UV} \left[ \phi^{}_{\rm B} \right] = \Gamma^{}_{\rm EFT} \left[ \phi^{}_{\rm B} \right] \;,
\label{eq:matching-condition}
%     (2)
\end{eqnarray}
where the effective actions on both sides are understood as the functionals of the light background fields $\phi^{}_{\rm B}$. Some comments on the matching principle indicated in Eq.~(\ref{eq:matching-condition}) are helpful. 
\begin{itemize}
	\item The background field method will be always utilized in our discussions. See, e.g., Ref.~\cite{Abbott:1981ke}, for a pedagogical introduction to the background field method and earlier relevant works. In this framework, the generating functional of the Green functions in the UV theory reads
\begin{eqnarray}
Z^{}_{\rm UV} \left[ J^{}_\Phi, J^{}_\phi \right] = \int \mathcal{D} \Phi \mathcal{D} \phi \exp \left\{\rmI \int {\rm d}^d x \left( \mathcal{L}^{}_{\rm UV} \left[ \Phi, \phi \right] + J^{}_\Phi \Phi + J^{}_\phi \phi \right) \right\} \;,
\label{eq:gen-fun-corr}
%     (3)
\end{eqnarray}
where ${\cal L}^{}_{\rm UV}[\Phi, \phi]$ stands for the Lagrangian of the UV theory with a heavy field $\Phi$ and a light field $\phi$, while $J^{}_\Phi$ and $J^{}_\phi$ are the external sources for the heavy and light fields, respectively. Notice that we have denoted $d \equiv 4 - 2\varepsilon$ (with $\varepsilon \to 0$) as the spacetime dimension and will use the dimensional regularization of the divergent integrals throughout~\cite{tHooft:1972tcz, Bollini:1972ui, Ashmore:1972uj}. After splitting the relevant fields into the background and quantum fields, i.e., $\Phi = \Phi^{}_{\rm B} + \Phi^\prime$ and $\phi = \phi^{}_{\rm B} + \phi^\prime$, one needs to perform the path integrals only over the quantum fields $\Phi^\prime$ and $\phi^\prime$. The background fields $\Phi^{}_{\rm B}$ and $\phi^{}_{\rm B}$ fulfill the classical equations of motion (EOMs), i.e.,
\begin{eqnarray}
	\frac{\delta \mathcal{L}^{}_{\rm UV}}{\delta \Phi} \left[ \Phi^{}_{\rm B}, \phi^{}_{\rm B} \right] + J^{}_\Phi &=& 0 \;,
	\nonumber
	\\
	\frac{\delta \mathcal{L}^{}_{\rm UV}}{\delta \phi} \left[ \Phi^{}_{\rm B}, \phi^{}_{\rm B} \right] + J^{}_\phi &=& 0 \;,
\label{eq:cl-eom}
%     (4)
\end{eqnarray}
in the presence of external sources. Around the background fields, we expand the generating functional $Z^{}_{\rm UV}[J^{}_\Phi, J^{}_\phi]$ up to the second order of the quantum fields and integrate over them to obtain
\begin{eqnarray}
Z^{}_{\rm UV} \left[ J^{}_\Phi, J^{}_\phi \right] 
\propto  \exp \left\{ \rmI \int {\rm d}^d x \left( \mathcal{L}^{}_{\rm UV} \left[ \Phi^{}_{\rm B}, \phi^{}_{\rm B} \right] + J^{}_\Phi \Phi^{}_{\rm B} + J^{}_\phi \phi^{}_{\rm B} \right) \right\} \times \left( \det \mathcal{Q}^{}_{\rm UV} \right)^{-c^{}_{\rm s}} \;,
\label{eq:approx-gf}
%     (5)
\end{eqnarray}
where 
\begin{eqnarray}
\mathcal{Q}^{}_{\rm UV}  \equiv \left(\begin{matrix} \displaystyle  -\frac{\delta^2 \mathcal{L}^{}_{\rm UV} }{\delta \Phi^2 } \left[ \Phi^{}_{\rm B}, \phi^{}_{\rm B} \right] &\hspace{0.3cm} \displaystyle -\frac{\delta^2 \mathcal{L}^{}_{\rm UV} }{\delta \Phi \delta \phi } \left[ \Phi^{}_{\rm B}, \phi^{}_{\rm B} \right] \\
\displaystyle -\frac{\delta^2 \mathcal{L}^{}_{\rm UV} }{\delta \phi \delta \Phi } \left[ \Phi^{}_{\rm B}, \phi^{}_{\rm B} \right] &\hspace{0.3cm} \displaystyle -\frac{\delta^2 \mathcal{L}^{}_{\rm UV} }{ \delta \phi^2 } \left[ \Phi^{}_{\rm B}, \phi^{}_{\rm B} \right] \end{matrix} \right) \equiv \left(\begin{matrix} \Delta^{}_{\Phi} & X^{}_{\Phi \phi} \\ X^{}_{\phi \Phi} & \Delta^{}_{\phi} \end{matrix}\right) \;,
\label{eq:ex-quadratic}
%     (6)
\end{eqnarray}
and $c^{}_{\rm s}$ accounts for the spin-statistic factor and the number of degrees of freedom, namely, $c^{}_{\rm s} = 1/2$ for real bosonic fields, $c^{}_{\rm s} = 1$ for complex bosonic fields, and $c^{}_{\rm s} = -1$ (or $-1/2$) for Dirac (or Majorana) fermionic fields. In the following, we set $c^{}_{\rm s} = 1/2$ and take account of the extra degrees of freedom by including the complex-conjugate fields [cf. Eq.~(\ref{eq:varphii})] and the minus sign for fermions. Then, the 1LPI effective action $\Gamma^{}_{\rm L, UV}[\phi^{}_{\rm B}]$ is defined as the Legendre transform of the generating functional of connected Green functions with $J^{}_\Phi = 0$, namely,
\begin{eqnarray}
\Gamma^{}_{\rm L, UV} \left[ \phi^{}_{\rm B} \right] &\equiv& \int {\rm d}^d x \mathcal{L}^{}_{\rm UV} \left[ \Phi^{}_{\rm c} \left[ \phi^{}_{\rm B} \right], \phi^{}_{\rm B} \right] + \frac{\rmI}{2} \ln \det \mathcal{Q}^{}_{\rm UV} \left[ \Phi^{}_{\rm c} \left[ \phi^{}_{\rm B} \right], \phi^{}_{\rm B} \right] \;,
\label{eq:1LPI-action}
%     (7)
\end{eqnarray}
where the classical heavy field $\Phi^{}_{\rm c}[\phi^{}_{\rm B}] \equiv \Phi^{}_{\rm B}[J^{}_\Phi = 0, J^{}_\phi]$ is determined by the EOM
\begin{eqnarray}
\left.\frac{\delta \mathcal{L}^{}_{\rm UV} \left[ \Phi, \phi \right] }{ \delta \Phi } \right|^{}_{\Phi = \Phi^{}_{\rm c} \left[ \phi^{}_{\rm B} \right] ,\, \phi = \phi^{}_{\rm B}} = 0 \;.
\label{eq:cl-eom-Phi}
%     (8)
\end{eqnarray}
with the vanishing external source $J^{}_\Phi = 0$ for the heavy field. Since the classical heavy field $\Phi^{}_{\rm c}[\phi^{}_{\rm B}]$ determined by Eq.~(\ref{eq:cl-eom-Phi}) is in general non-local, one can expand it to a given order of $1/M$ (where $M$ represents the mass scale of heavy fields) and thus denote the localized one by $\widehat{\Phi}^{}_{\rm c}[\phi^{}_{\rm B}]$. The final result of the 1LPI effective action $\Gamma^{}_{\rm L, UV}[\phi^{}_{\rm B}]$ is given by Eq.~(\ref{eq:1LPI-action}) with $\Phi^{}_{\rm c}[\phi^{}_{\rm B}]$ everywhere replaced by $\widehat{\Phi}^{}_{\rm c}[\phi^{}_{\rm B}]$.

\item On the other hand, we can similarly derive the generating functional of Green functions in the corresponding low-energy EFT, for which the Lagrangian will be denoted as ${\cal L}^{}_{\rm EFT}[\phi]$. At the one-loop level, the generating functional for the EFT reads
\begin{eqnarray}
Z^{}_{\rm EFT} \left[ J^{}_\phi \right] \propto \exp \left\{ \rmI \int {\rm d}^d x \left( \mathcal{L}^{\rm tree}_{\rm EFT} \left[ \phi^{}_{\rm B} \right] + \mathcal{L}^{\rm 1-loop}_{\rm EFT} \left[ \phi^{}_{\rm B} \right] + J^{}_\phi \phi^{}_{\rm B} \right) \right\} \times \left( \det \mathcal{Q}^{}_{\rm EFT} \right)^{-1/2} \;,
\label{eq:gen-fun-corr-eft}
%     (9)
\end{eqnarray}
where the EFT Lagrangian has been split into the tree-level one ${\cal L}^{\rm tree}_{\rm EFT}[\phi]$ and the one-loop-level one ${\cal L}^{\rm 1-loop}_{\rm EFT}[\phi]$, and we have defined
\begin{eqnarray}
	\mathcal{Q}^{}_{\rm EFT} \equiv -\frac{\delta^2 \mathcal{L}^{\rm tree}_{\rm EFT} }{\delta \phi^2} \left[ \phi^{}_{\rm B} \right] \;,
\label{eq:ex-quadratic-eft}
%     (10)
\end{eqnarray}
arising from the tree-level EFT Lagrangian. From the generating functional in Eq.~(\ref{eq:gen-fun-corr-eft}), one can get the 1PI effective action $\Gamma^{}_{\rm EFT}[\phi^{}_{\rm B}]$ up to the one-loop level
\begin{eqnarray}
\Gamma^{}_{\rm EFT} \left[ \phi^{}_{\rm B} \right] = \int {\rm d}^d x \left( \mathcal{L}^{\rm tree}_{\rm EFT} \left[ \phi^{}_{\rm B} \right] + \mathcal{L}^{\rm 1-loop}_{\rm EFT} \left[ \phi^{}_{\rm B} \right] \right) + \frac{\rmI}{2} \ln \det \mathcal{Q}^{}_{\rm EFT} \;,
\label{eq:1LPI-action-eft}
%     (11)
\end{eqnarray}
where the field multiplets [cf. Eq.~(\ref{eq:varphii})] have been introduced, as in the UV theory, to account for the number of degrees of freedom for the light fields in the EFT.

\item At the energy scale $\mu = M$, where $M$ can be identified as the masses of heavy fields in the UV theory, the matching condition in Eq.~(\ref{eq:matching-condition}) can be spelled out by using the 1LPI effective action $\Gamma^{}_{\rm L, UV}[\phi^{}_{\rm B}]$ in Eq.~(\ref{eq:1LPI-action}) and the 1PI effective action $\Gamma^{}_{\rm EFT}[\phi^{}_{\rm B}]$ in Eq.~(\ref{eq:1LPI-action-eft}). Following the arguments given in Refs.~\cite{Cohen:2020qvb, Fuentes-Martin:2020udw, Zhang:2021jdf}, we can complete the tree-level matching by implementing the EOM in Eq.~(\ref{eq:cl-eom-Phi}) for the classical heavy field $\Phi^{}_{\rm c}[\phi^{}_{\rm B}]$, i.e.,
\begin{eqnarray}
	\mathcal{L}^{\rm tree}_{\rm EFT} \left[ \phi^{}_{\rm B} \right] = \mathcal{L}^{}_{\rm UV} \left[ \widehat{\Phi}^{}_{\rm c} \left[ \phi^{}_{\rm B} \right], \phi^{}_{\rm B} \right] \;,
\label{eq:tree-matching}
%     (12)
\end{eqnarray}
where $\Phi^{}_{\rm c}[\phi^{}_{\rm B}]$ has been replaced by its localized counterpart. Moreover, the one-loop effective action of the EFT is given by
\begin{eqnarray}
 	\int {\rm d}^d x \mathcal{L}^{\rm 1-loop}_{\rm EFT} \left[ \phi^{}_{\rm B} \right] = \left. \Gamma^{\rm 1-loop}_{\rm L, UV} \left[ \phi^{}_{\rm B} \right] \right|^{}_{\rm hard} = \left. \frac{\rmI}{2} \ln \det \mathcal{Q}^{}_{\rm UV} \left[ \widehat{\Phi}^{}_{\rm c} \left[ \phi^{}_{\rm B} \right], \phi^{}_{\rm B} \right] \right|^{}_{\rm hard} \;,
 \label{eq:loop-matching3}
%     (13) 
\end{eqnarray}
where the subscripts ``hard" refer to the contributions from the hard-momentum region as the loop integrals are treated by the  expansion-by-regions techniques~\cite{Beneke:1997zp, Smirnov:2002pj, Jantzen:2011nz}. In practice, the result in Eq.~(\ref{eq:loop-matching3}) can be evaluated as~\cite{Cohen:2020fcu}
\begin{eqnarray}
 \int {\rm d}^d x \mathcal{L}^{\rm 1-loop}_{\rm EFT} \left[ \phi \right] = \left. \frac{\rmI}{2} {\rm STr} \ln \left( - \bm{K} \right) \right|^{}_{\rm hard} - \left. \frac{\rmI}{2} \sum^{\infty}_{n=1} \frac{1}{n} {\rm STr} \left[ \left( \bm{K}^{-1} \bm{X} \right)^n \right] \right|^{}_{\rm hard} \;,
\label{eq:loop-matching4}
%     (14) 
\end{eqnarray}
where the supertrace ``STr'' is the generalization of the trace over both the internal degrees of freedom and the functional space (with the fermionic blocks assigned a minus sign). Notice that the inverse-propagator part $\bm{K}$ and the interaction part $\bm{X}$ stem from the explicit calculation of $\ln \det \mathcal{Q}^{}_{\rm UV}$ on the rightmost side of Eq.~(\ref{eq:loop-matching3}), where $\bm{K}^{-1} \bm{X} \sim M^{-1}$ has been taken into account~\cite{Fuentes-Martin:2020udw}. As indicated in Eq.~(\ref{eq:loop-matching4}), there are both {\it log-type} and {\it power-type} supertraces, corresponding to the first and second terms in Eq.~(\ref{eq:loop-matching4}), respectively. The log-type supertrace is universal and receives contributions solely from heavy fields, leading to pure gauge-field operators. As we shall see later, the essential difference between the SEFT-I and the SEFT-II is that heavy fields are fermionic singlets in the former case while scalar triplets in the latter. Such a difference leads to extra dim-6 operators from the log-type supertraces in the SEFT-II.
\end{itemize}

Once the inverse-propagator part $\bm{K}$ and the interaction part $\bm{X}$ in the UV theory are obtained, the functional supertraces in Eq.~(\ref{eq:loop-matching4}) can be evaluated by means of the covariant derivative expansion (CDE) method~\cite{Henning:2014wua, Cohen:2019btp, Fuentes-Martin:2020udw}. Currently two {\sf Mathematica} packages, i.e., {\sf SuperTracer}~\cite{Fuentes-Martin:2020udw} and {\sf STrEAM}~\cite{Cohen:2020fcu}, making use of the CDE method to evaluate all functional supertraces, are publicly available. In the present work, we utilize the package {\sf SuperTracer} (specifically, the first version of {\sf SuperTracer}) in our calculations.

\section{The Type-II Seesaw Model}\label{sec:model}

Unlike the type-I seesaw model~\cite{Minkowski:1977sc, Yanagida:1979as, GellMann:1980vs, Glashow:1979nm, Mohapatra:1979ia}, where three right-handed neutrino singlets are introduced to the SM, the type-II seesaw model~\cite{Konetschny:1977bn, Magg:1980ut, Schechter:1980gr, Cheng:1980qt, Mohapatra:1980yp, Lazarides:1980nt} extends the SM by an ${\rm SU}(2)^{}_{\rm L}$ Higgs triplet with a hypercharge $Y = -1$. Now that the UV theory in question is the type-II seesaw model, the gauge-invariant Lagrangian for the UV theory is given by
\begin{eqnarray}
\mathcal{L}^{}_{\rm UV} = \mathcal{L}^{}_{\rm SM} + \frac{1}{2} \text{Tr} \left[\left(D_\mu\Delta\right)^\dag \left(D^\mu\Delta\right)\right] 
- V\left(H,\Delta\right) - \frac{1}{2} \left[ \overline{\ell^{}_{\rm L}} Y_\Delta\Delta \epsilon \ell_{\rm L}^{\rm c} + \text{h.c.} \right]  \; ,
\label{eq:Lagrangian}
%     (15)
\end{eqnarray}
where $\ell^{}_{\rm L}$ denotes the left-handed lepton doublet, $\epsilon = \rmI \sigma^2$ is the two-dimensional antisymmetric tensor with $\epsilon^{}_{12} = -\epsilon^{}_{21} = 1$, where the indices of this tensor are referring to the weak isospin space, and $\ell^{\rm c}_{\rm L} \equiv {\sf C} \overline{\ell^{}_{\rm L}}^{\rm T}$ with ${\sf C}\equiv \rmI \gamma^2 \gamma^0$ being the charge-conjugation matrix is defined, and the Higgs triplet $\Phi \equiv (\Phi^{}_1, \Phi^{}_2, \Phi^{}_3)$ has been cast in the matrix form $\Delta \equiv \sigma^
I \Phi^{}_I$ with $\sigma^I$ (for $I = 1, 2, 3$) being the Pauli matrices. In Eq.~(\ref{eq:Lagrangian}), the SM Lagrangian ${\cal L}^{}_{\rm SM}$ reads
%without the quadratic and quartic terms of the Higgs doublet $H$ reads
\begin{eqnarray}
\mathcal{L}^{}_{\rm SM} &=& -\frac{1}{4} G^A_{\mu\nu} G^{A\mu\nu} -\frac{1}{4} W^I_{\mu\nu} W^{I\mu\nu} -\frac{1}{4} B^{}_{\mu\nu} B^{\mu\nu} + \left( D^{}_\mu H \right)^\dagger \left( D^\mu H \right) - m^2H^\dag H - \lambda\left(H^\dag H\right)^2
\nonumber
\\
&&+ \sum^{}_f \overline{f} \rmI \slashed{D} f - \left( \overline{Q^{}_{\rm L}} Y^{}_{\rm u} \widetilde{H} U^{}_{\rm R} + \overline{Q^{}_{\rm L}} Y^{}_{\rm d} H D^{}_{\rm R} + \overline{\ell^{}_{\rm L}} Y^{}_l H E^{}_{\rm R} + {\rm h.c.} \right) \;,
\label{eq:Lagrangian-SM}
%     (16)
\end{eqnarray}
where $f=Q^{}_{\rm L}, U^{}_{\rm R}, D^{}_{\rm R}, \ell^{}_{\rm L}, E^{}_{\rm R}$ refer to the SM fermionic doublets and singlets, $H$ is the SM Higgs doublet with the hypercharge $Y = 1/2$, $\widetilde{H}$ is defined as $\widetilde{H} \equiv \epsilon H^*$, and the covariant derivative $D^{}_\mu \equiv \partial^{}_\mu - \rmI g^{}_1 Y B^{}_\mu - \rmI g^{}_2 T^I W^I_\mu - \rmI g^{}_s T^A G^A_\mu$ has been defined as usual. For the fields in the fundamental representation of ${\rm SU}(2)^{}_{\rm L}$, we have $T^I = \sigma^I/2$ (for $I = 1, 2, 3$). For the Higgs triplet $\Phi$ in the adjoint representation, we should take the representation matrices $(T^I)^{}_{JK} = -{\rm i} \epsilon^{IJK}$ (for $I, J, K = 1, 2, 3$), where $\epsilon^{IJK}$ is the totally antisymmetric Levi-Civita tensor. Note that $T^A = \lambda^A/2$ has been defined with $\lambda^A$ being the Gell-Mann matrices (for $A = 1, 2, \cdots, 8$), where the Latin letters $A, B, C$ refer to the adjoint representation of the ${\rm SU}(3)^{}_{\rm c}$ group while $I, J, K$ to that of the ${\rm SU}(2)^{}_{\rm L}$ group. In addition, the scalar potential $V(H, \Delta)$ without the quadratic and quartic terms of the Higgs doublet $H$ in Eq.~(\ref{eq:Lagrangian}) can be written as
\begin{eqnarray}
\begin{aligned}
		V\left(H,\Delta\right)=& \frac{1}{2}M_\Delta^2\text{Tr}\left(\Delta^\dag\Delta\right)-
		\left(\lambda_\Delta M_\Delta H^{\rm T} \epsilon \Delta H+\text{h.c.}\right) + \frac{\lambda^{}_1}{4} \left[\text{Tr}\left(\Delta^\dag\Delta\right)\right]^2+
		\frac{\lambda^{}_2}{4} \text{Det}\left(\Delta^\dag\Delta\right)
		\\
		& + \frac{\lambda^{}_3}{2} \left(H^\dag H\right) \text{Tr}\left(\Delta^\dag\Delta\right) + \frac{\lambda^{}_4}{2} \left(H^\dag \sigma^I H\right) \text{Tr}\left(\Delta^\dag \sigma^I \Delta\right) \; .
\end{aligned}
\label{eq:VDelH}
%     (17)
\end{eqnarray}
%where the first two terms should be combined with Eq.~(\ref{eq:Lagrangian-SM}) to reproduce the full SM Lagrangian, and the other terms are new in the type-II seesaw model. 
It is worthwhile to mention that both the trilinear doublet-triplet term in $V(H, \Delta)$ and the Yukawa interaction term in Eq.~(\ref{eq:Lagrangian}) are indispensable for the generation of tiny Majorana neutrino masses. Without loss of generality, the coupling constant $\lambda^{}_\Delta$ is taken to be a real parameter in this work. 

As is well known, after integrating out the Higgs triplet at the tree level, one can derive the dim-5 Weinberg operator. In this section, we shall derive the full tree-level Lagrangian for the SEFT-II by following the functional approach introduced in Sec.~\ref{sec:framework}. For later convenience, we rewrite the Lagrangian by expressing the Higgs triplet in the adjoint representation $\Phi$ instead of the matrix form $\Delta$. More explicitly, we have 
\begin{equation}
\begin{aligned}
\mathcal{L}^{}_{\text{UV}} \supset & \left(D_\mu\Phi\right)^\dag \left(D^\mu\Phi\right) - M_\Delta^2 \left(\Phi^\dag\Phi\right)  +{\widehat{\lambda}}^\dag_{I}{\Phi}_I+{\widehat{Y}}^\dag_I \Phi_I + \Phi^\dag_I \widehat{Y}_I+\Phi^\dag_I \widehat{\lambda}_{I}- \left(\lambda_1+\frac{\lambda_2}{4}\right)\left(\Phi^\dag\Phi\right)^2
\\ 
& + \frac{\lambda_2}{4}\left(\Phi^\dag T^I\Phi\right)\left(\Phi^\dag T^I\Phi\right)-
\lambda^{}_3 \left(H^\dag H\right) \left(\Phi^\dag\Phi\right) - \lambda^{}_4 \left(H^\dag \sigma^I H\right) \left(\Phi^\dag T^I \Phi\right) \; ,
\end{aligned}
\label{eq:typeIILag}
%     (18)
\end{equation}
where the repeated index ``$I$" should be summed over $I = 1, 2, 3$, the representation matrices $(T^I)_{JK} = -\rmI \epsilon^{IJK}$ are implied, and the relevant terms $\widehat{\lambda}^{}_I$ and $\widehat{Y}^{}_I$ are defined as
\begin{eqnarray}
	\widehat{\lambda}^{}_I \equiv -\lambda^{}_\Delta M^{}_\Delta H^\dagger \sigma^I \widetilde{H} \; , \qquad \widehat{Y}^{}_I \equiv \frac{1}{2} \overline{\ell^{\rm c}_{\rm L}} \rmI \sigma^2 \sigma^I Y^\dagger_\Delta \ell^{}_{\rm L} \; .
\label{eq:lambdaYhat}
%     (19)
\end{eqnarray}
Therefore, it is straightforward to derive the EOM of the Higgs triplet from Eq.~(\ref{eq:typeIILag}), i.e.,
\begin{equation}
\begin{aligned} 
D^{}_\mu D^\mu \Phi = & - M_\Delta^2 \Phi + \widehat{\lambda} + \widehat{Y} - \left(2\lambda^{}_1 + \frac{\lambda^{}_2}{2} \right)
\left(\Phi^\dag\Phi\right) \Phi + \frac{\lambda^{}_2}{2} \left(\Phi^\dag T^I \Phi\right) \left(T^I \Phi \right) \\ &
- \lambda^{}_3 \left(H^\dag H\right)\Phi - \lambda^{}_4 \left(H^\dag \sigma^I H\right) \left(T^I \Phi\right) \; ,
\end{aligned}
\label{eq:typeIIeom}
%     (20)
\end{equation}
from which one can determine the classical triplet Higgs field $\Phi^{}_{\rm c}$. In order to find out the operators in the SEFT-II up to dim-6, we need to localize $\Phi^{}_{\rm c}$ by expanding it to a given order of $M^{-1}_\Delta$ and maintain only the terms up to ${\cal O}(M^{-4}_\Delta)$. Hence we finally arrive at
\begin{eqnarray}
	\widehat{\Phi}^{}_{I, \rm c} = \left[ \frac{1}{M^2_\Delta} \delta^{}_{IJ} - \frac{1}{M^4_\Delta} \left( D^2_{IJ} + \lambda^{}_3 H^\dagger H \delta^{}_{IJ} + \lambda^{}_4 H^\dag \sigma^K H T^K_{IJ} \right) \right] \left( \widehat{\lambda}^{}_J + \widehat{Y}^{}_J \right) + {\cal O}(M^{-6}_\Delta) \; .
\label{eq:classicalPhi}
%     (21)
\end{eqnarray}
% \begin{eqnarray}
% 	\widehat{\Phi}^{}_{I, \rm c} = \frac{1}{M^2_\Delta} \left( \widehat{\lambda}^{}_I + \widehat{Y}^{}_I \right) + {\cal O}(M^{-4}_\Delta) = \frac{1}{M_\Delta^2} \left( \frac{1}{2} \overline{\ell^{\rm c}_{\rm L}} \epsilon \sigma^I Y_\Delta^\dag \ell^{}_{\rm L} -
% 	\lambda_\Delta M^{}_\Delta H^\dag \sigma^I \widetilde{H} \right) + {\cal O}\left(M_\Delta^{-4}\right) \; ,
% \label{eq:classicalPhi}
% %     (21)
% \end{eqnarray}
%where the definitions in Eq.~(\ref{eq:lambdaYhat}) have been used. 

Then inserting the classical triplet Higgs field in Eq.~(\ref{eq:classicalPhi}) into Eq.~(\ref{eq:typeIILag}), we can immediately obtain the effective operators up to dim-6 in the SEFT-II at the tree level. But before doing so, to make this procedure more transparent, we recast Eq.~\eqref{eq:typeIILag} into
\begin{eqnarray}
    \mathcal{L}^{}_{\rm UV} &\supset&  \Phi^\dag \left[ - D^2 \Phi - M^2_\Delta \Phi + \widehat{\lambda} + \widehat{Y}  - \left(2\lambda^{}_1 + \frac{\lambda^{}_2}{2} \right) \left(\Phi^\dag\Phi\right) \Phi + \frac{\lambda^{}_2}{2} \left(\Phi^\dag T^I \Phi\right) \left(T^I \Phi \right) \right.
    \nonumber
    \\
    && - \left. \lambda^{}_3 \left(H^\dag H\right)\Phi - \lambda^{}_4 \left(H^\dag \sigma^I H\right) \left(T^I \Phi\right) \vphantom{\frac{1}{1}}\right] + \left(\lambda_1+\frac{\lambda_2}{4}\right)\left(\Phi^\dag\Phi\right)^2 - \frac{\lambda_2}{4}\left(\Phi^\dag T^I\Phi\right)\left(\Phi^\dag T^I\Phi\right)
    \nonumber
    \\
    && + {\widehat{\lambda}}^\dag_{I}{\Phi}_I + {\widehat{Y}}^\dag_I \Phi_I \;.
    \label{eq:typeIILag-I}
\end{eqnarray}
The terms in the square brackets satisfy the EOM of the Higgs triplet field, thus they will be vanishing when the classical triplet Higgs field $\widehat{\Phi}^{}_{\rm c}$ in Eq.~\eqref{eq:classicalPhi} is inserted. Moreover, the two $\Phi^4$ terms at the end of the second line in Eq.~\eqref{eq:typeIILag-I} will only result in the operators of mass dimension higher  than six and can be safely neglected. Therefore, the tree-level operators induced by integrating out the heavy Higgs triplet field are all included in 
\begin{eqnarray}
    \mathcal{L}^{}_{\rm SEFT-II} \supset \left( \widehat{\lambda}^\dag_I + \widehat{Y}^\dag_I \right) \left[ \frac{1}{M^2_\Delta} \delta^{}_{IJ} - \frac{1}{M^4_\Delta} \left( D^2_{IJ} + \lambda^{}_3 H^\dagger H \delta^{}_{IJ} + \lambda^{}_4 H^\dag \sigma^K H T^K_{IJ} \right) \right] \left( \widehat{\lambda}^{}_J + \widehat{Y}^{}_J \right) \;,
    \label{eq:tree-level-matching}
\end{eqnarray}
where the classical triplet Higgs field given in Eq.~\eqref{eq:classicalPhi} has been applied. Some interesting observations can be made.

\begin{itemize}
    \item Notice that the mass dimension of $\widehat{Y}^{}_I$ in Eq.~(\ref{eq:lambdaYhat}) is three, contributed from two lepton doublets, whereas $\widehat{\lambda}^{}_I$ contains only two Higgs doublets that are totally of mass dimension two. Thus one dimension-four operator comes out, i.e.,
    \begin{eqnarray}
	    \frac{1}{M^2_\Delta} \left(\widehat{\lambda}^\dagger \widehat{\lambda}\right) =  \lambda^2_\Delta \left(
	    H^\dagger \sigma^I \widetilde{H} \right)^\dagger \left(
    	H^\dagger \sigma^I \widetilde{H} \right) = 2  \lambda^2_\Delta \left(H^\dagger H\right)^2 \; ,
        \label{eq:4Higgs}
        %     (26)
    \end{eqnarray}
    where the identity $(\sigma^I)_{ab} (\sigma^I)^{}_{cd} = 2 \delta^{}_{ad} \delta^{}_{bc} - \delta^{}_{ab} \delta^{}_{cd}$ has been utilized. The four-Higgs operator in Eq.~(\ref{eq:4Higgs}) leads to the tree-level threshold effect on the quartic Higgs coupling, namely,
    \begin{eqnarray}
	    \lambda^{}_{\rm eff} = \lambda - 2 \lambda^2_\Delta  \; .
	    \label{eq:lambdaeff}
        %     (27)
    \end{eqnarray}
    The minus sign on the right-hand side of Eq.~(\ref{eq:lambdaeff}) is crucially important, since the threshold shift in the quartic coupling of the Higgs doublet may help rescue the electroweak vacuum from instability, similar to the scenario of the SM with an extra scalar singlet considered in Ref.~\cite{Elias-Miro:2012eoi}. In addition, the vacuum expectation value (vev) of the SM Higgs field will be modified to $v= \sqrt{-m^2/(\lambda - 2\lambda^2_\Delta)} \simeq 246$ GeV once the effects of the dimension-four operator in Eq.~\eqref{eq:4Higgs} are considered.
    
	\item It is easy to see that the cross terms of $\widehat{\lambda}$ and $\widehat{Y}$ give rise to the unique dim-5 operator~\cite{Weinberg:1979sa}
	\begin{equation}
		\sum_i \frac{ C^{(5)}_i O^{(5)}_i}{\Lambda} = \frac{1}{M_\Delta^2} \left({\widehat{Y}}^\dag\widehat{\lambda} + {\widehat{\lambda}}^\dag \widehat{Y}\right)
		= - \frac{\left(Y^{}_\Delta \right)^{}_{\alpha\beta} \lambda^{}_\Delta}{M^{}_\Delta} \left[\overline{\ell^{}_{\alpha\rm L}} \widetilde{H} {\widetilde{H}}^{\rm T} \ell_{\beta\rm L}^{\rm c} \right] + \text{h.c.} \; .
		\label{eq:dim5}
		%     (22)
	\end{equation}
Hence the dim-5 operator and its Wilson coefficient are given by
\begin{eqnarray}
	O^{(5)}_{\alpha \beta} = \overline{\ell^{}_{\alpha \rm L}} \widetilde{H} {\widetilde{H}}^{\rm T} \ell_{\beta \rm L}^{\rm c} \; , \quad C^{(5)}_{\alpha \beta} = -\lambda^{}_\Delta \left(Y^{}_\Delta\right)^{}_{\alpha \beta} \; ,
	\label{eq:C5O5}
	%     (23)
\end{eqnarray}
with $\alpha, \beta$ running over the lepton flavors $e, \mu, \tau$, and the cutoff energy scale is identified as $\Lambda = M^{}_\Delta$. After the SM Higgs field acquires its vev, i.e., $\langle H \rangle = v/\sqrt{2}$ and the SM gauge symmetry is spontaneously broken, the dim-5 operator gives rise to a Majorana neutrino mass term with the effective neutrino mass matrix $M^{}_\nu = \lambda^{}_\Delta Y^{}_\Delta  v^2/M^{}_\Delta$. Therefore, the smallness of neutrino masses can be attributed to the heaviness of the Higgs triplet (i.e., $M^{}_\Delta \gg v$), manifesting the spirit of seesaw mechanisms.

\item One can find that there are three dim-6 operators at the tree level. Among them one is the four-fermion operator stemming from the $\widehat{Y}^2/M^2_\Delta$ term, namely
\begin{eqnarray}
\frac{C^{(6)}_{4\ell} O^{(6)}_{4\ell}}{\Lambda^2} = \frac{1}{M_\Delta^2} \left({\widehat{Y}}^\dag\widehat{Y}\right) = \frac{1}{4M_\Delta^2}
	\left(\overline{\ell^{}_{\rm L}} Y^{}_\Delta \sigma^I \epsilon \ell_{\rm L}^{\rm c} \right)^\dag \left(\overline{\ell^{}_{\rm L}} Y^{}_\Delta \sigma^I \epsilon \ell_{\rm L}^{\rm c} \right) \; ,
	\label{eq:fourlO}
%     (24)
\end{eqnarray}
from which we can identify the dim-6 operators with lepton flavors specified
\begin{eqnarray}
	O^{(6)}_{4\ell, \alpha \beta \gamma \delta} = \left(\overline{\ell^{}_{\alpha \rm L}} \sigma^I \epsilon \ell_{\beta \rm L}^{\rm c} \right)^\dagger \left(\overline{\ell^{}_{\gamma \rm L}} \sigma^I \epsilon \ell_{\delta \rm L}^{\rm c} \right) \; , \quad C^{(6)}_{4\ell, \alpha \beta \gamma \delta} = (Y^{}_\Delta)^{*}_{\alpha \beta} (Y^{}_\Delta)^{}_{\gamma \delta}/4 \; .
	\label{eq:fourlOflavor}
%     (25)
\end{eqnarray}
Another two operators result from the $\widehat{\lambda}^2/M^4_\Delta$ terms. To be specific, we have
\begin{eqnarray}
	\frac{C^{(6)}_{D^2 H^4} O^{(6)}_{D^2 H^4}}{\Lambda^2} = \frac{1}{M_\Delta^4}\left(D_\mu\widehat{\lambda}\right)^\dag\left(D^\mu\widehat{\lambda}\right)
	= \frac{ \lambda^2_\Delta }{M_\Delta^2} \left[D^{}_\mu \left(H^\dag \sigma^I \widetilde{H}\right)\right]^\dag \left[ D^\mu \left(H^\dag \sigma^I \widetilde{H}\right) \right] \; ,
	\label{eq:D2H4Oper}
%     (28)
\end{eqnarray}
and
\begin{eqnarray}
		\frac{C^{(6)}_{H^6} O^{(6)}_{H^6}}{\Lambda^2} &=&  - \frac{\lambda^{}_3}{M_\Delta^4} \left(H^\dag H\right) \left({\widehat{\lambda}}^\dag \widehat{\lambda}\right)
		- \frac{\lambda^{}_4}{M_\Delta^4} \left(H^\dag \sigma^I H\right) \left({\widehat{\lambda}}^\dag T^I \widehat{\lambda}\right) \nonumber \\ &=& -\frac{2(\lambda^{}_3 - \lambda^{}_4) \lambda^2_\Delta }{M^2_\Delta} \left(H^\dag H\right)^3 \; , 
		\label{eq:H6Oper}
%     (29)
\end{eqnarray}
from which one can easily extract those two dim-6 operators and the corresponding Wilson coefficients. More explicitly, from Eq.~\eqref{eq:H6Oper}, we get $O^{(6)}_{H^6} = \left(H^\dag H\right)^3$ and $C^{(6)}_{H^6} = -2(\lambda^{}_3 - \lambda^{}_4) \lambda^2_\Delta $, and
\begin{eqnarray}
O^{(6)}_{D^2 H^4} &=& \left[D^{}_\mu \left(H^\dag \sigma^I \widetilde{H}\right)\right]^\dag \left[ D^\mu \left(H^\dag \sigma^I \widetilde{H}\right) \right] \; , \quad C^{(6)}_{D^2 H^4} =  \lambda^2_\Delta 
\; , 
\label{eq:O6tree}
%     (30)
\end{eqnarray}
from Eq.~(\ref{eq:D2H4Oper}). 
\end{itemize}

The aforementioned operators in the SEFT-II have already been partially or completely derived in the previous works~\cite{Chao:2006ye, Schmidt:2007nq, Abada:2007ux}.\footnote{It is worthwhile to mention that the dim-6 operators $O^{(6)}_{H^6}$ and $O^{(6)}_{D^2H^4}$ from the tree-level matching have not been discussed in Refs.~\cite{Chao:2006ye, Schmidt:2007nq} but in Ref.~\cite{Abada:2007ux}. We have compared our results in Eqs. \eqref{eq:4Higgs}-\eqref{eq:O6tree} with those obtained in Ref.~\cite{Abada:2007ux} and found an excellent agreement.} Now we should convert those operators and the corresponding Wilson coefficients into those in the Warsaw basis by applying the Fierz transformation, integration by parts and the EOM of the Higgs doublet at the tree level. As a result, the tree-level matching leads to
\begin{eqnarray}
    \mathcal{L}^{\rm tree}_{\rm SEFT-II} &=& \mathcal{L}^{}_{\rm SM} +2\lambda^2_\Delta\left(1+\frac{2m^2}{M_\Delta^2}\right) \left(H^\dagger H\right)^2 + \left(\frac{C^{(5)}_{\alpha\beta}}{M^{}_\Delta}  O^{(5)}_{\alpha \beta} + {\rm h.c.} \right) + \frac{\left(C^{\rm tree}_{\ell\ell}\right)^{}_{\alpha\beta\gamma\delta}}{M^2_\Delta} O^{\alpha\beta\gamma\delta}_{\ell\ell}  
    \nonumber
    \\
    && + \frac{C^{\rm tree}_{H}}{M^2_\Delta} O^{}_{H} + \frac{C^{\rm tree}_{H\square}}{M^2_\Delta} O^{}_{H\square}  + \frac{C^{\rm tree}_{HD}}{M^2_\Delta} O^{}_{HD} + \frac{1}{M^2_\Delta} \left[ \left(C^{\rm tree}_{eH}\right)^{}_{\alpha\beta} O^{\alpha\beta}_{eH}  + \left(C^{\rm tree}_{uH}\right)^{}_{\alpha\beta} O^{\alpha\beta}_{uH} \right.
    \nonumber
    \\
    && \left.  + \left(C^{\rm tree}_{dH}\right)^{}_{\alpha\beta} O^{\alpha\beta}_{dH} + {\rm h.c.} \right] \;, \quad
    \label{eq:lagrangian-tree}
\end{eqnarray}
where the dim-6 operators $O^{}_H$, $O^{}_{H\square}$, $O_{HD}^{}$, $O_{eH}^{}$, $O_{uH}^{}$, $O_{dH}^{}$ and $O^{}_{\ell\ell}$ are in the Warsaw basis, and their specific forms and the associated Wilson coefficients are respectively given by 
\begin{eqnarray}
    && O^{}_H = \left( H^\dagger H \right)^3 \;,\quad O^{}_{H\square} = \left( H^\dagger H \right) \square \left( H^\dagger H\right) \;,\quad
    O_{HD}=\left( H^\dag D_\mu^{} H \right)^\dag \left( H^\dag D^\mu H \right) \;,
    \nonumber
    \\
    && O^{\alpha\beta}_{eH} = \left( H^\dagger H \right) \left( \BLelli \REi H \right) \;,\quad O^{\alpha\beta}_{uH} = \left( H^\dagger H \right) \left( \BLQi \RUi \widetilde{H} \right) \;,\quad O^{\alpha\beta}_{dH} = \left( H^\dagger H \right) \left( \BLQi \RDi H \right) 
    \nonumber
    \\
    && O^{\alpha\beta\gamma\delta}_{\ell\ell} = \left( \BLelli \gamma^\mu \Lelli \right)\left( \BLelli[\gamma] \gamma^{}_\mu \Lelli[\delta] \right) \;.
    \label{eq:tree-operators}
\end{eqnarray}
and
\begin{eqnarray}\label{eq:tree-operators}
&& C_{H}^{\rm tree} = 2(4\lambda-\lambda_3+\lambda_4)\lambda_\Delta^2-16\lambda_\Delta^4 \;,\quad C_{H\square}^{{\rm tree}} = 2\lambda_\Delta^2 \;,\quad
C_{HD}^{\rm tree}=4\lambda_\Delta^2 \;,
\nonumber
\\
&&  \left(C_{eH}^{_{\rm tree}}\right)_{\alpha\beta} = 2\lambda_\Delta^2\left(Y_l\right)_{\alpha\beta} \;,\quad
\left(C_{uH}^{\rm tree}\right)_{\alpha\beta} = 2\lambda_\Delta^2\left(Y_{\rm u}\right)_{\alpha\beta} \;,\quad  \left(C_{dH}^{\rm tree}\right)_{\alpha\beta} = 2\lambda_\Delta^2\left(Y_{\rm d}\right)_{\alpha\beta}
\nonumber
\\
&& \left(C_{\ell \ell}^{\rm tree}\right)_{\alpha\beta\gamma\delta} = \frac{1}{4} \left(Y^{}_\Delta\right)_{\alpha\gamma} \left(Y_\Delta^\dag\right)_{\beta\delta} \;.
\label{eq:tree-WC}
\end{eqnarray}

Thus far we have accomplished the tree-level matching of the type-II seesaw model onto the SMEFT. In addition to the unique dim-5 Weinberg operator, there are seven dim-6 operators in the Warsaw basis. The decoupling of the Higgs triplet also leads to the threshold shift in the quartic Higgs coupling $\lambda \to \lambda - 2 \lambda^2_\Delta \left( 1 + 2m^2/M^2_\Delta \right) $ below the decoupling scale (i.e., $\mu < M^{}_\Delta$). As a direct consequence, the vev of the SM Higgs field will be modified.

\section{One-loop Matching}\label{sec:oneloop}

As has been explained in Sec.~\ref{sec:framework}, one should calculate the matrices $\bm{K}$ and $\bm{X}$ in the UV theory in order to accomplish the one-loop matching. In the type-II seesaw model, the inverse-propagator part can be wirtten as
\begin{eqnarray}
\bm{K}^{}_\Phi = P^2 - M^2_\Delta \; ,
\label{eq:KPhi}
%     (31)
\end{eqnarray}
where $P^{}_\mu \equiv \rmI D^{}_\mu$ with $D^{}_\mu$ being the covariant derivative. Note that the gauge boson fields in the covariant derivative should be replaced by their background fields, and the inverse-propagator $\bm{K}^{}_\Phi$ is universal for the heavy Higgs triplet $\Phi^{}_I$. 

Furthermore, we define the field multiplets $\varphi^{}_i$ that are relevant for one-loop matching in the type-II seesaw model as below
\begin{eqnarray}
	\varphi^{}_i \in \{ \varphi^{}_\Phi,\; \varphi^{}_\ell,\; \varphi^{}_E,\; \varphi^{}_Q,\; \varphi^{}_U,\; \varphi^{}_D,\; \varphi^{}_H,\; \varphi^{}_W,\; \varphi^{}_B \} \;,
\end{eqnarray}
where
\begin{eqnarray}
	\varphi^{}_\Phi = \left( \begin{matrix}
		\Phi \cr \Phi^*	\end{matrix}\right) \;,\quad \varphi^{}_F = \left(\begin{matrix} F \\ F^{\rm c} \end{matrix} \right) \;,\quad \varphi^{}_H = \left( \begin{matrix} H \\  H^* \end{matrix} \right) \;,\quad \varphi^{}_V = V \;,
\label{eq:varphii}
%     (33)
\end{eqnarray}
with $F =\ell, E, Q, U, D$ and $V=W, B$. To extract the interaction matrix $\bm{X}$, one needs to first figure out the fluctuation operator in Eq.~(\ref{eq:ex-quadratic}) and then follow the procedure from Eq.~(\ref{eq:loop-matching3}) to Eq.~(\ref{eq:loop-matching4}), see, e.g., Ref.~\cite{Cohen:2020fcu}, for more details. In the following, we list all the relevant $X$ terms arising from the type-II seesaw model

\begin{itemize}
	\item $X^{}_{HH}$
	\begin{equation}
		\begin{aligned} 
			X^{}_{HH}&=\left(\begin{matrix} X^{}_{11}
				&X^{}_{12}\\ X^{}_{21} & X^{}_{22} \\\end{matrix}\right)
		\end{aligned}
	\end{equation}
	with
	\begin{equation}
		\begin{aligned} 
			X^{}_{11} &=   m^2+2\lambda(\left|H\right|^2+HH^\dag) + \lambda^{}_3 (\Phi^\dag \Phi) +
			\lambda^{}_4 \sigma^I (\Phi^\dag T^I \Phi ) \\
			X^{}_{12} &=2 \lambda H H^{\rm T} +\lambda^{}_\Delta M^{}_\Delta\Phi_I^\dag
			\left(\sigma^I \epsilon\right) \\
			X^{}_{21} &= 2\lambda H^\ast H^\dag - \lambda^{}_\Delta M^{}_\Delta \left(\epsilon\sigma^I\right) \Phi^{}_I \\
			X^{}_{22} &= m^2+2\lambda(\left|H\right|^2+H^\ast H^{\rm T})
			+\lambda^{}_3(\Phi^\dag \Phi)+\lambda^{}_4 \sigma^{I\rm T}(\Phi^\dag T^I \Phi) 
		\end{aligned}
	\end{equation}
	\item $X^{}_{\ell\Phi}$
	\begin{equation}
		\begin{aligned} 
			X_{\ell\Phi}^I=\left(\begin{matrix} \displaystyle \frac{Y^{}_\Delta}{2}\left(\sigma^I \epsilon {\ell}_{\rm L}^{\rm c}\right)
				&0\\0& \displaystyle \frac{Y_\Delta^\ast}{2}\left(\epsilon\sigma^I {\ell}_{\rm L}\right)\\\end{matrix}\right)
		\end{aligned}
	\end{equation}
	\item $X^{}_{\Phi \ell}$
	\begin{equation}
		\begin{aligned} 
			X_{\Phi \ell}^I=\left(\begin{matrix} \displaystyle \frac{Y_\Delta^\dag}{2}\left({\ell}_L^{\rm T} {\sf C}\epsilon\sigma^I\right)
				&0\\0& \displaystyle \frac{Y_\Delta^{\rm T}}{2}\left(\overline{{\ell}_{\rm L}}\sigma^I\epsilon\right)\\\end{matrix}\right) 
		\end{aligned}
	\end{equation}
	\item $ X_{W\Phi}^{}$
	\begin{equation}
		\begin{aligned} 
			X_{W\Phi}^I&=\rmI g_2^{}\left(\begin{matrix} \left(D^{}_\mu\Phi\right)^\dag T^I
				&-\left(D^{}_\mu\Phi\right)^{\rm T }T^{I\ast}\\\end{matrix}\right)+g_2^{}\left(\begin{matrix}-\Phi^\dag T^I
				&\Phi^{\rm T}T^{I\ast}\\\end{matrix}\right)\rmI D^{}_\mu
		\end{aligned}
	\end{equation}  
	\item $ X_{\Phi W}^{}$
	\begin{equation}
		\begin{aligned} 
			X_{\Phi W}^I &=2\rmI g_2^{}\left(\begin{matrix}-T^I(D^{}_\mu\Phi)\\T^{I\ast}\left(D^{}_\mu\Phi\right)^\ast\\\end{matrix}\right)+g_2^{}\left(\begin{matrix}-T^I\Phi\\T^{I\ast}\Phi^\ast\\\end{matrix}\right)\rmI D^{}_\mu
		\end{aligned}
	\end{equation}
	\item $X^{}_{\Phi B}$
	\begin{equation}
		\begin{aligned} 
			X^{}_{\Phi B}&=2\rmI g_1^{} \left(\begin{matrix}\left(D^{}_\mu\Phi\right)\\{-\left(D^{}_\mu\Phi\right)}^\ast\\\end{matrix}\right)
			+g_1^{}\left(\begin{matrix}\Phi\\-\Phi^\ast\\\end{matrix}\right)\rmI D^{}_\mu
		\end{aligned}
	\end{equation}
	\item $ X^{}_{B\Phi}$
	\begin{equation}
		\begin{aligned} 
			X^{}_{B\Phi}&=\rmI g_1^{} \left(-\begin{matrix}\left(D^{}_\mu\Phi\right)^\dag&\left(D^{}_\mu\Phi\right)^{\rm T}\\\end{matrix}\right)
			+g_1^{}\left(\begin{matrix}\Phi^\dag&-\Phi^{\rm T}\\\end{matrix}\right)\rmI D^{}_\mu
		\end{aligned}
	\end{equation}
	\item $X^{}_{H\Phi}$
	\begin{equation}
		\begin{aligned} 
			X_{H\Phi}^I=\left(\begin{matrix}X^{}_{11}&X^{}_{12}\\X^{}_{21}&X^{}_{22}\\\end{matrix}\right) 
		\end{aligned}
	\end{equation} 
	with
	\begin{equation}
		\begin{aligned} 
			X^{}_{11}=&\lambda^{}_3 H\Phi^\dag + \lambda^{}_4 \left(\sigma^J H\right)
			\left(\Phi^\dag T^J \right)\\
			X^{}_{12}=&2\lambda^{}_\Delta M^{}_\Delta \left(\sigma^I \epsilon H^\ast\right)
			+\lambda^{}_3 H\Phi^{\rm T} + \lambda^{}_4 \left(\sigma^J H\right)\left(\Phi^{\rm T}T^{J \ast}\right)\\ 
			X^{}_{21}=&-2\lambda^{}_\Delta M^{}_\Delta
			\left(\epsilon\sigma^I H\right) + \lambda^{}_3 H^\ast\Phi^\dag + \lambda^{}_4 \left(\sigma^{J \ast} H^\ast\right)
			\left(\Phi^\dag T^J \right)\\
			X^{}_{22}=& \lambda^{}_3 H^\ast \Phi^{\rm T} + \lambda^{}_4 \left(\sigma^{J \ast} H^\ast\right) \left(\Phi^{\rm T}T^{J \ast}\right)
		\end{aligned}
	\end{equation}
	\item $X^{}_{\Phi H}$
	\begin{equation}
		\begin{aligned} 
			X_{\Phi H}^I =\left(\begin{matrix}X^{}_{11} & X^{}_{12} \\ X^{}_{21} & X^{}_{22} \\ \end{matrix}\right) 
		\end{aligned}
	\end{equation} 
	with
	\begin{equation}
		\begin{aligned}
			X^{}_{11}=&\lambda^{}_3 H^\dag \Phi + \lambda^{}_4 \left(H^\dag\sigma^J\right)
			\left(T^J \Phi\right) \\
			X^{}_{12}=& 2\lambda^{}_\Delta M^{}_\Delta \left(H^\dag \sigma^I \epsilon \right)
			+ \lambda^{}_3 H^{\rm T}\Phi + \lambda^{}_4 \left(H^{\rm T}\sigma^{J \ast} \right) \left(T^J\Phi\right)\\
			X^{}_{21} = &-2\lambda^{}_\Delta M^{}_\Delta
			\left(H^{\rm T}\epsilon\sigma^I \right) + \lambda^{}_3 H^\dag \Phi^\ast + \lambda^{}_4 \left(H^\dag\sigma^J\right)
			\left(T^{J \ast} \Phi^\ast\right)\\
			X^{}_{22}=&\lambda^{}_3 H^{\rm T} \Phi^\ast + \lambda^{}_4\left(H^{\rm T}\sigma^{J \ast} \right)
			\left(T^{J \ast} \Phi^\ast\right)
		\end{aligned}
	\end{equation}
	\item $ X^{}_{\Phi\Phi}$
	\begin{equation}
		\begin{aligned} 
			X_{\Phi\Phi}^{JK}=\left(\begin{matrix}X^{}_{11} & X^{}_{12} \\ X^{}_{21} & X^{}_{22} \\ \end{matrix}\right) 
		\end{aligned}
	\end{equation}
	with
	\begin{equation}
		\begin{aligned} 
			X^{}_{11}=& \left( 2 \lambda^{}_1 + \frac{\lambda^{}_2}{2}\right) \left(|\Phi|^2 + \Phi \Phi^\dag \right)
			-\frac{\lambda^{}_2}{2} \left[T^{I} \left(\Phi^\dag T^I \Phi\right) + \left(T^I\Phi\right) \left(\Phi^\dag T^I \right) \right]\\
			& + \lambda^{}_3 \left(H^\dag H\right) + \lambda^{}_4 \left(H^\dag\sigma^I H\right)T^I \\ 
			X^{}_{12}=&\left(2\lambda^{}_1 + \frac{\lambda^{}_2}{2} \right) \Phi \Phi^{\rm T} - \frac{\lambda^{}_2}{2}
			\left(T^I \Phi\right) \left(\Phi^{\rm T}T^I\right) \\
			X^{}_{21}= & \left(2\lambda^{}_1 + \frac{\lambda^{}_2}{2}\right) \Phi^\ast\Phi^\dag - \frac{\lambda^{}_2}{2}
			\left(T^I \Phi^\ast\right) \left(\Phi^\dag T^I \right) \\
			X^{}_{22}=& \left(2\lambda^{}_1 + \frac{\lambda^{}_2}{2}\right) \left(\Phi^\dag\Phi +\Phi^\ast\Phi^{\rm T}\right)
			-\frac{\lambda^{}_2}{2}\left[T^I \left(\Phi^\dag T^I \Phi\right)+\left(T^{I \ast}\Phi^\ast\right) \left(\Phi^{\rm T}T^I \right) \right]\\
			& +\lambda^{}_3\left(H^\dag H\right) +\lambda^{}_4\left(H^\dag\sigma^I H\right)T^I
		\end{aligned}
	\end{equation}
	\item $X^{}_{HB}$
	\begin{equation}
		\begin{aligned} 
			X^{}_{HB}& =\rmI g_1^{}\left(\begin{matrix}-
				D^{}_\nu H\\\left(D^{}_\nu H\right)^\ast\\\end{matrix}\right)
			+ \frac{g_1^{}}{2}\left(\begin{matrix}-H\\H^\ast\\\end{matrix}\right)\rmI D^{}_\nu
		\end{aligned}
		\label{eq:X-SM-1}
	\end{equation}
	\item $X^{}_{BH}$
	\begin{equation}
		\begin{aligned} 
			X^{}_{BH} & =\frac{\rmI g_1^{}}{2}\left(\begin{matrix}\left(D^{}_\nu H\right)^\dag&
				-\left(D^{}_\nu H\right)^{\rm T}\\\end{matrix}\right)+\frac{g_1^{}}{2}\left(\begin{matrix}
				-H^\dag&H^{\rm T}\\\end{matrix}\right)\rmI D^{}_\nu
		\end{aligned}
	\end{equation}
	\item $X_{HW}^{}$
	\begin{equation}
		\begin{aligned} 
			X_{HW}^I& =\rmI g_2^{}\left(\begin{matrix}-\sigma^I D^{}_\nu H\\\sigma^{I\ast}\left(D^{}_\nu H\right)^\ast\\\end{matrix}\right)
			+\frac{g_2^{}}{2}\left(\begin{matrix}-\sigma^I H\\ \sigma^{I\ast} H^\ast \\ \end{matrix}\right)\rmI D^{}_\nu
		\end{aligned}
	\end{equation}
	\item $X_{WH}^{}$
	\begin{equation}
		\begin{aligned} 
			X_{WH}^I &=\frac{\rmI g_2^{}}{2}\left(\begin{matrix}\left(D^{}_\nu H\right)^\dag\sigma^I&
				-\left(D^{}_\nu H\right)^{\rm T} \sigma^{I\ast}\\ \end{matrix}\right)
  	+\frac{g_2^{}}{2}\left(\begin{matrix}-H^\dag\sigma^I & H^{\rm T}\sigma^{I\ast}\\\end{matrix}\right)\rmI D^{}_\nu
		\end{aligned}
		\label{eq:X-SM-2}
	\end{equation}
\end{itemize}
In the $X$ terms, the heavy Higgs triplet field $\Phi^{}_I$ has to be replaced by the solution to its EOM, namely, the right-hand side of Eq.~(\ref{eq:classicalPhi}). The corresponding $X$ terms for the SM interactions can be found in the Appendix B of Ref.~\cite{Cohen:2020fcu}, whereas the $X$ terms for the interactions between the Higgs boson and gauge bosons should take the forms given in Eqs.~(\ref{eq:X-SM-1})-(\ref{eq:X-SM-2})  due to different conventions used in the package {\sf SuperTracer}~\cite{Fuentes-Martin:2020udw} and in Ref.~\cite{Cohen:2020fcu}. In addition, different conventions of the quartic Higgs coupling in the literature should be noted.

With the above information, it is now ready to evaluate the supertraces in Eq.~(\ref{eq:loop-matching4}) and then to derive the one-loop-level operators by using the package {\sf SuperTracer}. Then, the generated operators from {\sf SuperTracer} can be further converted into the independent operators in the Warsaw basis of the SMEFT~\cite{Grzadkowski:2010es}. However, in order to cross-check our results via diagrammatic approach, we first convert these operators into those in Green's basis~\cite{Jiang:2018pbd,Gherardi:2020det,Chala:2021cgt} by utilizing the algebraic, Fierz identities and integration by parts. In the following subsection, we shall discuss the one-loop threshold corrections to the renormalizable terms existing in the SM Lagrangian, and clarify how to implement the EOMs to remove the redundant dim-6 operators in a consistent way. The concept of field redefinitions and its impact on the Wilson coefficients of dim-6 operators in the Warsaw basis will be emphasized~\cite{Criado:2018sdb}.

\subsection{Threshold corrections}\label{sec:4.1}

Generally, the one-loop matching can result in threshold corrections to the renormalizable terms already existing in the SM~\cite{Wells:2017vla}, i.e.,
\begin{eqnarray}\label{eq:correction-GB}
	\delta \mathcal{L} &=& \delta Z^{}_{G} G^A_{\mu\nu} G^{A\mu\nu} + \delta Z^{}_W W^I_{\mu\nu} W^{I\mu\nu} + \delta Z^{}_B B^{}_{\mu\nu} B^{\mu\nu}
	\nonumber
	\\
	&&+ \sum^{}_f  \overline{f} \delta Z^{}_f \rmI \slashed{D} f + \left( \overline{Q^{}_{\rm}} \delta Y^{}_{\rm u} \widetilde{H} U^{}_{\rm R} + \overline{Q^{}_{\rm L}} \delta Y^{}_{\rm d} H D^{}_{\rm R} + \overline{\ell^{}_{\rm L}} \delta Y^{}_l H E^{}_{\rm R} + {\rm h.c.} \right)
	\nonumber
	\\
	&& + \delta Z^{}_H \left( D^{}_\mu H \right)^\dagger \left( D^\mu H \right) + \delta m^2 H^\dagger H + \delta \lambda \left( H^\dagger H \right)^2 \;.
\end{eqnarray}
where $f=Q^{}_{\rm L}, U^{}_{\rm R}, D^{}_{\rm R}, \ell^{}_{\rm L}, E^{}_{\rm R}$. For a given UV model, not all the above terms are induced, which is of course dependent on the interactions of the heavy fields. In the type-II seesaw model, the one-loop matching leads to
\begin{eqnarray}\label{eq:renormalizable}
	 \left( 4\pi \right)^2 \delta Z^{\rm G}_B  &=&  -\frac{g_1^2 L_\Delta}{4} \;,
	\nonumber
	\\
    \left( 4\pi \right)^2 \delta Z^{\rm G}_W  &=&  -\frac{g_2^2 L_\Delta}{6}\;,
	\nonumber
	\\
	\left( 4\pi \right)^2 \delta Z^{\rm G}_H &=&  6\lambda^2_\Delta + \frac{6m^2}{M_\Delta^2}\lambda_\Delta^2\left(5+2L^{}_\Delta\right) \;,  
	\nonumber
	\\
	\left( 4\pi \right)^2 \left( \delta Z^{\rm G}_\ell \right)^{}_{\alpha\beta} &=& \frac{3}{4}\left(1+2L^{}_{\Delta}\right)\left(Y^{{}}_\Delta Y_\Delta^\dag\right)_{\alpha\beta} \;, 
	\nonumber
	\\
	\left( 4\pi \right)^2 \left( \delta m^2 \right)^{\rm G} &=& 3 \left[ M_\Delta^2 \left(4\lambda^2_\Delta + \lambda_3 \right)+ 4m^2\lambda_\Delta^2 +\frac{4m^4}{M_\Delta^2}\lambda_\Delta^2 \right] \left(1+L^{}_\Delta\right) \;,
	\nonumber
	\\
\left( 4\pi \right)^2 \delta \lambda^{\rm G} &=& \frac{1}{2} \left( 3\lambda_3^2 + 2 \lambda_4^2 \right) L_\Delta^{} + 2 \lambda^2_\Delta \left[ \left( 20 \lambda + 8\lambda^{}_1 + \lambda^{}_2 \right) \left( 1 + L^{}_\Delta \right) - 2\lambda^{}_3 \left( 5 + 2 L^{}_\Delta \right) \right.
\nonumber
\\
&&  + \left.  4\lambda^{}_4 \left( 3 + 2L^{}_\Delta \right) - 20\lambda^2_\Delta \left( 2 + L^{}_\Delta \right) \right]  + \frac{4m^2}{M^2_\Delta} \lambda^2_\Delta \left[ 20\lambda \left( 1 + L^{}_\Delta \right) - \lambda^{}_3 \left( 8 + 5L^{}_\Delta \right)  \right.
\nonumber
\\
&&  + \left. 2\lambda^{}_4 \left( 4 + 3 L^{}_\Delta \right) - 20 \lambda^2_\Delta \left( 3 + 2L^{}_\Delta \right) \right] \;,
\end{eqnarray}
with
\begin{eqnarray}\label{eq:log-def}
	L^{}_{\Delta} \equiv \ln \left( \frac{\mu^2}{M^2_{\Delta}} \right) + \frac{1}{\varepsilon} - \gamma^{}_{\rm E} + \ln \left( 4\pi \right)\; ,
\end{eqnarray}
up to $\mathcal{O}\left( M^{-2}_\Delta \right)$ in the Green's basis, where the other terms in Eq.~(\ref{eq:correction-GB}) do not appear, the superscript ``G" signifies the results in the Green's basis, and $\gamma^{}_{\rm E}$ in Eq. (\ref{eq:log-def}) is the Euler constant. It is worth pointing out that the divergences in $L^{}_\Delta$ come from the hard part of loop integrals and usually consist of both the UV and infrared (IR) divergences in the dimensional regularization. The UV divergences can be absorbed by the renormalization constants in the UV model but with heavy fields replaced with their classical EOMs, such as Eq.~(\ref{eq:classicalPhi}) in the type-II seesaw model, while the IR divergences can be regarded as part of the counterterms of the EFT to cancel corresponding UV divergences of the EFT~\cite{Dittmaier:2021fls}. Here, one can simply remove the $1/\varepsilon -\gamma^{}_{\rm E} + \ln \left( 4\pi \right)$ terms in $L^{}_\Delta$ to get the renormalized couplings or Wilson coefficients in the $\overline{\rm MS}$ scheme~\cite{Bilenky:1993bt,Dittmaier:2021fls}.

In addition, there also exist one-loop corrections to the Wilson coefficient of the dimension-five operator $\left(\delta C^{(5)} \right)^{\rm G}_{\alpha\beta} O^{(5)}_{\alpha\beta}$ in Eq.~(\ref{eq:C5O5}), namely,
\begin{eqnarray}
	\left( 4\pi \right)^2 \left( \delta C^{(5)} \right)^{\rm G}_{\alpha\beta} = \lambda_\Delta\left(1+L_\Delta\right)\left[ \left(2\lambda_3-4\lambda_4-8\lambda_1-\lambda_2\right) (Y^{}_\Delta)_{\alpha \beta} + (Y^{}_lY_l^\dag Y^{}_\Delta)_{\alpha \beta} + (Y^{}_\Delta Y^{\ast}_l Y_l^{\rm T} )_{\alpha \beta}\right] \;. \nonumber \\
\end{eqnarray}
This result will be important for us to find the threshold correction to the Wilson coefficient of the Weinberg operator in the Warsaw basis.

With the help of the EOMs of relevant fields, we can obtain the operators together with the Wilson coefficients in the Warsaw basis from those in the Green's basis. This is normally done in the diagrammatic calculations for one-loop matching, as in Refs.~\cite{Gherardi:2020det,Zhang:2021tsq}. However, as has been stressed in Ref.~\cite{Criado:2018sdb}, the implementation of the EOMs of fields to get rid of the operator redundancy, even including higher-order corrections, may not completely reproduce the correct effective Lagrangian. Without repeating the general arguments in Ref.~\cite{Criado:2018sdb}, we just explain how to consistently use the EOMs and demonstrate that it is necessary to supplement the results with additional terms from field redefinitions.
\begin{itemize}
	\item First, it is important to clarify which EOMs of fields should be implemented. Apart from the existing renormalizable terms in the tree-level effective Lagrangian in the Green's basis, we have to take into account the threshold corrections in Eq.~\eqref{eq:correction-GB} from one-loop matching. The overall Lagrangian in the Green's basis with both tree-level and one-loop corrections should be used to derive the lowest-order EOMs in the SEFT-II, namely,
\begin{eqnarray}\label{eq:EOM-II}
	\rmI \slashed{D} E^{}_{\rm R} &=& H^\dagger Y^\dagger_l \ell^{}_{\rm L} \;,
	\nonumber
	\\
	\rmI \slashed{D} \ell^{}_{\rm L} &=& \left( 1 - \delta Z^{\rm G}_\ell \right) Y^{}_l H E^{}_{\rm R} \;,
	\nonumber
	\\
	D^\nu B^{}_{\mu \nu} &=& \frac{g^{}_1}{2} \left[ \left( 1 + \delta Z^{\rm G}_H + 4 \delta Z^{\rm G}_B \right) H^\dagger \rmI \Dlr H + 2  \left( 1 + 4 \delta Z^{\rm G}_B \right) \sum^{}_f Y(f) \overline{f} \gamma^{}_\mu f \right] \;,
	\nonumber
	\\
	\left( D^\nu W_{\mu \nu} \right)^I &=& \frac{g^{}_2}{2} \left[ \left( 1 + \delta Z^{\rm G}_H + 4 \delta Z^{\rm G}_W \right) H^\dagger \rmI \Dilr H + \left( 1 + 4 \delta Z^{\rm G}_W \right)  \left( \overline{Q^{}_{\rm L}} \tau^I \gamma^{}_\mu Q^{}_{\rm L} + \overline{\ell^{}_{\rm L}} \tau^I \gamma^{}_\mu \ell^{}_{\rm L} \right) \right] \;,
	\nonumber
	\\
	\left( D^2 H \right)^a &=& - \left[ m^2 - \left( \delta m^2 \right)^{\rm G} - m^2 \delta Z^{\rm G}_H \right] H^a - 2 \left[ \left( \lambda - 2\lambda^2_\Delta \right) \left( 1 - \delta Z^{\rm G}_H \right) -\delta \lambda^{\rm G} \right] \left( H^\dagger H \right) H^a
	\nonumber
	\\
	&& - \left( 1 - \delta Z^{\rm G}_H \right) \left( \overline{E^{}_{\rm R}} Y^\dagger_l \ell^a_{\rm L} - \overline{D^{}_{\rm R}} Y^\dagger_{\rm d} Q^a_{\rm L} + \epsilon^{ab} \overline{Q^b_{\rm L}} Y^{}_{\rm u}  U^{}_{\rm R} \right) \;,
\end{eqnarray}
where $Y(f)$ is the hypercharge for the fermionic fields $f= Q^{}_{\rm L}, U^{}_{\rm R}, D^{}_{\rm R}, \ell^{}_{\rm L}, E^{}_{\rm R}$, the covariant derivatives are defined as $\Dlr \equiv D^{}_\mu - \Dl$ and $\Dilr \equiv \sigma^I D^{}_\mu - \Dl \sigma^I$ with $\Dl$ acting on the left, and the superscripts $I=1,2,3$ and $a,b=1,2$ are implied. Note that only the terms of $\mathcal{O} \left( M^0_\Delta \right)$ and $\mathcal{O} \left( M^2_\Delta \right)$ in $\delta Z^{\rm G}$, $\left(\delta m^2 \right)^{\rm G}$ and $\delta \lambda^{\rm G}$ are retained, since others will contribute higher-order corrections.

\item Second, the application of the lowest-order EOMs in Eq.~(\ref{eq:EOM-II}) can indeed remove the operator redundancy in the Green's basis, but cannot always give rise to the correct Wilson coefficients of dim-6 operators in the Warsaw basis, as indicated by the general discussions in Ref.~\cite{Criado:2018sdb}. Following the arguments in Ref.~\cite{Criado:2018sdb}, we now briefly explain the main idea. The action $S[\phi]$ of the effective field theory can be organized as a power series in $\zeta \equiv 1/\Lambda$, i.e.,
\begin{eqnarray}\label{eq:expansion-action}
S[\phi] = \sum^{2}_{n=-2} \zeta^n S^{}_n [\phi] \; .
\end{eqnarray}
In Eq.~(\ref{eq:expansion-action}), the summation on the right-hand side begins with $n = -2$ instead of $n = 0$, since the threshold correction to the quadratic coupling of the Higgs doublet (i.e., $m^2$) may be of order $\mathcal{O} \left(\Lambda^2 \right)$, as shown in Eq. \eqref{eq:renormalizable} in the type-II seesaw model under consideration. Furthermore, the summation ends with $n = 2$, as only the operators up to dim-6 will be taken into account. Therefore, to eliminate redundant dim-6 operators, one can perform the following perturbative field redefinition
\begin{eqnarray}\label{eq:local-transformation-1}
\phi^\alpha \to  \phi^\alpha - \zeta^2 f^\alpha_2 (\phi) \;,
\end{eqnarray}
at the order of $\mathcal{O}(\zeta^2)$, where $f^\alpha_2(\phi)$ denotes the functional corresponding to the redefinition of $\phi^\alpha$. Then the action $S[\phi]$ changes into $S^\prime[\phi]$, namely
\begin{eqnarray}\label{eq:expansion-action-2}
S^\prime [\phi] &=& S[\phi] - \zeta^2 f^\alpha_2 (\phi) \frac{\delta}{\delta \phi^\alpha}\sum^{0}_{n=-2} \zeta^n S^{}_n [\phi] + \frac{1}{2} \zeta^2 f^\alpha_2 (\phi) f^\beta_2 (\phi) \frac{\delta^2 S^{}_{-2} [\phi] }{\delta \phi^\alpha \delta \phi^\beta } \;,
\end{eqnarray}
where only the terms up to the order of $\mathcal{O} (\zeta^2)$ are retained. As one can see from Eq. \eqref{eq:expansion-action-2}, the field redefinition in Eq. \eqref{eq:local-transformation-1} can also modify $S^{}_k$ (for $k=0,1$) due to the existence of $S^{}_{-2}[\phi]$ and $S^{}_{-1}[\phi]$, which are absent in the scenarios considered in Ref.~\cite{Criado:2018sdb} but will be present in general. If we choose $f^\alpha_2(\phi)$ by requiring $f^\alpha_2 (\phi) \delta \mathcal{K}/\delta \phi^\alpha$ to be a redundant term in $S^{}_2 [\phi]$, where $\mathcal{K}$ can be any term in $\sum^{0}_{n=-2} \zeta^n S^{}_n$ but usually taken to be the kinetic term. In this way, the redundant operator in $S^{}_2 [\phi]$ can be eliminated by the second term in Eq.~\eqref{eq:expansion-action-2}, which is equivalent to substitute the EOMs in the action $S[\phi]$. However, the last term of order $\mathcal{O} (\zeta^2)$ in Eq.~\eqref{eq:expansion-action-2} cannot be reproduced by the utilization of the EOMs. Therefore, the application of the EOMs and the field redefinition are not equivalent in the presence of $S^{}_{-2}[\phi]$ in the action.

In the SEFT-II, we find that $S^{}_{-2}[H,H^\dagger] = \int d^4 x [3/(4\pi)^2]\cdot (4\lambda^2_\Delta + \lambda^{}_3)(1+L^{}_\Delta) H^\dagger H$ with $\zeta = 1/M^{}_\Delta$ can be obtained with the help of Eq.~\eqref{eq:renormalizable}, which is induced at the one-loop level. Thus we only need to take account of the field redefinition at the tree level to evaluate the last term in Eq.~\eqref{eq:expansion-action-2}. First of all, after the tree-level dim-6 operator $O^{}_{D^2H^4}$ in Eq.~\eqref{eq:O6tree} is decomposed into those in the Green's basis, we shall find that the operator
\begin{eqnarray} \label{eq:tree-redundancy}
O^{\prime \rm G}_{HD} &=& (H^\dagger H) (D^{}_\mu H)^\dagger(D^\mu H) 
\nonumber
\\
&=& \frac{1}{2} \left( H^\dagger H \right) \square \left( H^\dagger H \right) - \frac{1}{2} \left( H^\dagger H \right) \left( D^2 H \right)^\dagger H - \frac{1}{2} \left( H^\dagger H \right) H^\dagger \left( D^2 H \right) 
\end{eqnarray} 
with the tree-level Wilson coefficient $C^{\prime \rm G}_{HD}|^{}_{\rm tree} = 4\lambda^2_\Delta$ appears in the Green's basis, but it is absent in the Warsaw basis. Then, to remove this redundant operator $O^{\prime \rm G}_{HD}$ [more exactly the last two operators in Eq.~\eqref{eq:tree-redundancy}, as the first one appears in the Warsaw basis], one can redefine the Higgs doublet as $H \to H - [1/(2M^2_\Delta)] C^{\prime \rm G}_{HD}|^{}_{\rm tree} \cdot (H^\dagger H) H$, and likewise for $H^\dagger$, namely, $f^1_2 (\phi) = (1/2)\cdot C^{\prime \rm G}_{HD}|^{}_{\rm tree} \cdot (H^\dagger H) H$, $f^2_2 (\phi) = (1/2)\cdot C^{\prime \rm G}_{HD}|^{}_{\rm tree} \cdot (H^\dagger H) H^\dagger$, and $\mathcal{K} = \left(D^\mu H \right)^\dagger \left(D^{}_\mu H \right)$ with $\phi^1$ and $\phi^2$ referring to $H$ and $H^\dagger$, respectively. Therefore, in addition to the results obtained from the direct use of the EOMs in Eq.~(\ref{eq:EOM-II}), there will be an extra contribution 
\begin{eqnarray}\label{eq:second-derivative}
	\frac{1}{2} \zeta^2 f^\alpha_2(\phi) f^\beta_2(\phi) \frac{\delta^2 S^{}_{-2}[\phi] }{\delta \phi^\alpha \delta \phi^\beta} &=& \frac{1}{M^2_\Delta} \cdot \frac{1}{4} \left(C^{\prime \rm G}_{HD}|^{}_{\rm tree}\right)^2 \cdot \frac{3}{(4\pi)^2}  (4\lambda^2_\Delta + \lambda^{}_3)(1+L^{}_\Delta) \cdot (H^\dagger H)^6 \nonumber \\
	&=& \frac{12}{(4\pi)^2 M^2_\Delta} \lambda^4_\Delta (4\lambda^2_\Delta + \lambda^{}_3)(1+L^{}_\Delta) \cdot (H^\dagger H)^6 \; ,
\end{eqnarray}
where $\alpha,\beta=1,2$ and the repeated indices on the left-hand side should be summed. The one-loop contribution in Eq.~\eqref{eq:second-derivative} will be missed by simply applying the EOMs and must be added to the Wilson coefficient of the dim-6 operator $\Op^{}_H = \left(H^\dagger H \right)^6$ in the Warsaw basis.
\end{itemize}

Applying the EOMs in Eq.~(\ref{eq:EOM-II}) to the dim-6 operators in the Green's basis, one obtains one-loop corrections to the renormalizable terms in the Warsaw basis:
\begin{eqnarray}\label{eq:GtoW}
\delta Z^{}_B &=& \delta Z^{\rm G}_B \;,
\nonumber
\\
\delta Z^{}_W &=& \delta Z^{\rm G}_W \;,
\nonumber
\\
\delta Z^{}_H &=& \delta Z^{\rm G}_H \;,
\nonumber
\\
\delta Z^{}_\ell &=& \delta Z^{\rm G}_\ell \;,
\nonumber
\\
\delta m^2 &=& \left( \delta m^2 \right)^{\rm G} + C^{\rm G}_{DH} \frac{m^4}{M^2_\Delta} \;,
\nonumber
\\
\delta \lambda &=& \delta \lambda^{\rm G} + \left[ -\frac{g_2^2}{2} C^{\rm G}_{2W} + 2g^{}_2 C^{\rm G}_{WDH} + 4(\lambda  - 2 \lambda_\Delta^2) C^{\rm G}_{DH} + C^{\prime \rm G}_{HD} \right] \frac{m^2}{M^2_\Delta} \nonumber \\
&& - \left[ \left( \delta m^2 \right)^{\rm G} + m^2 \delta Z^{\rm G}_H \right] \frac{C^{\prime \rm G}_{HD}|^{}_{\rm tree}}{M^2_\Delta} \;,
\nonumber
\\
\left( \delta \Yl \right)^{}_{\alpha\beta} &=&  \Yli{\beta} C^{\rm G}_{DH} \frac{m^2}{M^2_\Delta}\;,
\nonumber
\\
\left( \delta \Yu \right)^{}_{\alpha\beta} &=& \Yui{\beta} C^{\rm G}_{DH} \frac{m^2}{M^2_\Delta} \;,
\nonumber
\\
\left( \delta \Yd \right)^{}_{\alpha\beta} &=& \Ydi{\beta} C^{\rm G}_{DH} \frac{m^2}{M^2_\Delta}\;,
\end{eqnarray}
where $C^{\rm G}_{DH}$, $C^{\rm G}_{2W}$, $C^{\rm G}_{WDH}$ and $C^{\prime \rm G}_{HD}$ denote the one-loop Wilson coefficients of the dim-6 operators $O^{\rm G}_{DH}$, $O^{\rm G}_{2W}$, $O^{\rm G}_{WDH}$ and $O^{\prime \rm G}_{HD}$ in the Green's basis, respectively. One can see that $\delta Y^{}_l$, $\delta Y^{}_{\rm u}$ and $\delta Y^{}_{\rm d}$ absent in the Green's basis now appear in the Warsaw basis. They are induced by the operator $O^{\rm G}_{DH}$ in the Green's basis after applying the EOM of $H$. Moreover, $O^{\rm G}_{DH}$ also gives an additional one-loop contribution to the Wilson coefficient of the dim-5 operator
\begin{eqnarray}
\left( \delta C^{(5)} \right)^{}_{\alpha\beta} &=& \left( \delta C^{(5)} \right)^{\rm G}_{\alpha\beta} - 2 C^{(5)}_{\alpha\beta} C^{\rm G}_{DH} \frac{m^2}{M^2_\Delta}\;,
\end{eqnarray}
but this contribution is of the order of $\mathcal{O}\left(M^{-3}_\Delta \right)$, since the dim-5 operator term itself is suppressed by $M^{-1}_\Delta$ in the Lagrangian, and will be omitted. Then the kinetic terms of gauge bosons $W^I_\mu$, $B^{}_\mu$, and the $\rm SU(2)^{}_L$ doublets $H$ and $\ell^{}_{\rm L}$ need to be normalized. For the gauge bosons, one needs to redefine the gauge boson fields and gauge couplings $g^{}_1$ and $g^{}_2$ simultaneously, namely
\begin{eqnarray}
    \begin{cases}
    B^{}_\mu \to \left( 1 + 2\delta Z^{}_B \right) B^{}_\mu 
    \\
    g^{}_1 \to g^{\rm eff}_1 = \left( 1 - 2\delta Z^{}_B \right) g^{}_1
    \end{cases} 
    \;,\qquad
    \begin{cases}
    W^I_\mu \to \left( 1 + 2\delta Z^{}_W \right) W^I_\mu 
    \\
    g^{}_2 \to g^{\rm eff}_2 = \left( 1 - 2\delta Z^{}_W \right) g^{}_2
    \end{cases}
    \;,
    \label{eq:gauge-boson}
\end{eqnarray}
to keep the canonical form of the covariant derivative $D^{}_\mu$. Thus the normalization of the kinetic terms of gauge bosons leads to the threshold corrections to the gauge couplings, and the effective gauge couplings are given by 
\begin{eqnarray}
    g^{\rm eff}_1 = \left( 1 + \frac{g_1^2 L^{}_{\Delta}}{2(4\pi)^2} \right) g^{}_1 \;,\quad  g^{\rm eff}_2 = \left( 1 + \frac{g_2^2 L^{}_{\Delta}}{3(4\pi)^2} \right) g^{}_2 \;.
    \label{eq:g-eff}
\end{eqnarray}
It is worth pointing out that in the type-I seesaw model, there are no threshold corrections to gauge couplings $g^{}_1$ and $g^{}_2$, thanks to the absence of gauge interactions for the singlet heavy right-handed neutrinos~\cite{Zhang:2021jdf}. The kinetic terms of Higgs and lepton doublets can be normalized by
\begin{eqnarray}\label{eq:shift}
H \to \left( 1 - \frac{1}{2} \delta Z^{}_H \right) H \;,\quad \ell^{}_{\rm L} \to \left( 1 - \frac{1}{2} \delta Z^{}_\ell \right) \ell^{}_{\rm L} \;,
\end{eqnarray}
which give additional one-loop contributions to the effective couplings in the EFT via the corresponding tree-level terms. From Eqs.~(\ref{eq:GtoW})-(\ref{eq:shift}), we obtain the effective couplings in the EFT
\begin{eqnarray}
m^2_{\rm eff} &=& m^2 \left( 1 - \delta Z^{}_H \right) - \delta m^2 
\nonumber
\\
&=& m^2-\frac{1}{(4\pi)^2} \left[ 3 M_\Delta^2 \left( 4 \lambda_\Delta^2 + \lambda^{}_3 \right) \left( 1 + L^{}_\Delta \right) + 6 m^2 \lambda^2_\Delta \left( 3 + 2 L^{}_\Delta \right) \vphantom{\frac{m^4}{M^2_\Delta}} + \frac{4 m^4}{M_\Delta^2}\lambda_\Delta^2 \left( 11 + 6L^{}_\Delta \right) \right] \;,
\nonumber
\\
\lambda^{}_{\rm eff} &=& \left[\lambda-2\lambda_\Delta^2\left(1+2\frac{m^2}
{M_\Delta^2} \right)\right]\left(1-2\delta Z_H\right)-\delta \lambda
\nonumber
\\
&=& \lambda - 2\lambda_\Delta^2 \left( 1 + \frac{2m^2}{M_\Delta^2} \right) + \frac{1}{\left(4\pi\right)^2} \left\{  - \frac{1}{2} \left( 3\lambda^2_3 + 2\lambda^2_4 \right) L^{}_\Delta + \left( g^4_2 - 20 \lambda^2_4 \right) \vphantom{\frac{\lambda^2_\Delta}{3}} \frac{m^2}{30M^2_\Delta}  \right.
\nonumber
\\
&& + \left[  3g^2_1 \left( 5+ 6L^{}_\Delta \right) + g^2_2 \left( 61 + 86L^{}_\Delta \right) - 24 \lambda\left( 59 + 34 L^{}_\Delta \right) - 48 \left( 8 \lambda^{}_1 + \lambda^{}_2 \right) \left( 1 + L^{}_\Delta \right) \right.
\nonumber
\\
&& + \left. 12 \lambda^{}_3 \left( 29 + 18 L^{}_\Delta \right) - 8 \lambda^{}_4 \left( 59 + 42L^{}_\Delta \right) \right] \frac{m^2 \lambda^2_\Delta}{6 M^2_\Delta} - 2\lambda^2_\Delta \left[ 2\lambda \left( 13 + 10L^{}_\Delta \right)  \right. 
\nonumber
\\
&& + \left. \left( 8\lambda^{}_1 + \lambda^{}_2 \right) \left( 1+ L^{}_\Delta \right) - 2\lambda^{}_3 \left( 8 + 5 L^{}_\Delta \right) + 4 \lambda^{}_4 \left( 3 + 2L^{}_\Delta \right) \right] + \frac{8\lambda^4_\Delta}{3} \left[ 3 \left( 19 + 11 L^{}_\Delta \right) \vphantom{\frac{\lambda^4_\Delta}{M^2_\Delta}} \right.
\nonumber
\\
&& + \left.\left. 20 \left( 13 + 6L^{}_\Delta \right) \frac{m^2}{M^2_\Delta} \right] \right\} \;,
\nonumber
\\
\left( Y^{\rm eff}_{l} \right)^{}_{\alpha\beta} &=&  \left( Y_l\right)^{}_{\alpha\beta} \left( 1 - \delta Z^{}_H/2 \right)-\frac{1}{2} \delta Z_{\ell}^{} Y_l^{} - \left(\delta Y_l\right)^{}_{\alpha\beta} 
\nonumber
\\
&=& \left( Y_l^{} \right)_{\alpha\beta} -\frac{1}{\left(4\pi\right)^2} \left\{ \left[ 3 + \frac{m^2}{M_\Delta^2} \left( 17 + 6 L^{}_\Delta \right) \right] \lambda_\Delta^2 \left( Y_l^{} \right)_{\alpha\beta} + \frac{3}{8}\left( Y_\Delta^{} Y_\Delta^\dag Y_l^{} \right) \left( 1 + 2L^{}_\Delta \right) \vphantom{\frac{\lambda^4_\Delta}{M^2_\Delta}} \right\} \;,
\nonumber 
\\
\left( Y^{\rm eff}_{\rm u} \right)^{}_{\alpha\beta} &=&  \left( \Yu\right)^{}_{\alpha\beta} \left( 1 - \delta Z^{}_H/2 \right) - \left(\delta \Yu\right)^{}_{\alpha\beta} 
\nonumber
\\
&=& \left( Y^{}_{\rm u} \right)_{\alpha\beta} - \frac{1}{(4\pi)^2} \left[ 3 + \frac{m^2}{M_\Delta^2} \left( 17 + 6 L^{}_\Delta \right) \right] \lambda_\Delta^2 \left(Y^{}_{\rm u}\right)_{\alpha\beta} \;,
\nonumber
\\
\left( Y^{\rm eff}_{\rm d} \right)^{}_{\alpha\beta} &=&  \left( \Yd\right)^{}_{\alpha\beta} \left( 1 - \delta Z^{}_H/2 \right) - \left(\delta \Yd\right)^{}_{\alpha\beta} 
\nonumber
\\
&=& \left( Y^{}_{\rm d} \right)_{\alpha\beta} - \frac{1}{(4\pi)^2} \left[ 3 + \frac{m^2}{M_\Delta^2} \left( 17 + 6 L^{}_\Delta \right) \right] \lambda_\Delta^2 \left(Y^{}_{\rm d}\right)_{\alpha\beta} \;,
\label{eq:threshold-corrections}
\end{eqnarray}
up to $\mathcal{O}\left( M^{-2}_\Delta \right)$. Similarly, the Wilson coefficient of the dim-5 operator is given by
\begin{eqnarray}
\left( C^{(5)}_{\rm eff} \right)^{}_{\alpha\beta} &=& \left[ C^{(5)} \left( 1 - \delta Z^{}_H \right) - \frac{1}{2} \delta Z^{}_\ell C^{(5)} - \frac{1}{2} C^{(5)} \delta Z^{\rm T}_\ell + \delta C^{(5)} \right]^{}_{\alpha\beta} 
\nonumber
\\
&=& - \lambda_\Delta \left( Y^{}_\Delta \right)_{\alpha\beta} + \frac{\lambda_\Delta}{(4\pi)^2} 
\left\{ 6 \lambda^2_\Delta \left(Y^{}_\Delta\right)_{\alpha\beta} + \frac{3}{4} \left( 1 + 2L^{}_{\Delta} \right) \left( Y^{}_\Delta Y^{\dag}_\Delta Y^{}_\Delta \right)_{\alpha\beta} + \left( 1 + L_\Delta \right) \right.
\nonumber
\\
&& \times \left. \left[ \left( 2\lambda_3 - 4\lambda_4 - 8\lambda_1 - \lambda_2 \right) \left( Y_\Delta \right)^{}_{\alpha\beta} + \left( Y^{}_lY_l^\dag Y^{}_\Delta \right)_{\alpha \beta} + \left( Y^{}_\Delta Y^{\ast}_l Y_l^{\rm T} \right)_{\alpha \beta} \right]  \vphantom{\frac{\lambda^2}{4}}\right\} \;.
\label{eq:WC-dim-5}
\end{eqnarray} 
Those in Eqs.~\eqref{eq:g-eff}, \eqref{eq:threshold-corrections} and \eqref{eq:WC-dim-5} are the complete one-loop matching results, which can be used together with the two-loop renormalization-group equations (RGEs) of relevant Wilson coefficients in the SEFT-II.

\subsection{Dimension-six operators}

In the type-II seesaw model, the dim-6 operators $O^{}_H$, $O^{}_{H\square}$, $O_{HD}^{}$, $O_{eH}^{}$, $O_{uH}^{}$, $O_{dH}^{}$ and $O^{}_{\ell\ell}$ in the Warsaw basis have already appeared after integrating out the heavy triplet scalar at the tree level, as shown in Sec.~\ref{sec:model}, and the tree-level contributions to their Wilson coefficients are shown in Eq.~\eqref{eq:tree-WC}. Those tree-level Wilson coefficients in Eq.~\eqref{eq:tree-WC} result in extra one-loop contributions to the total Wilson coefficients of $O^{}_H$, $O^{}_{H\square}$, $O_{HD}^{}$, $O_{eH}^{}$, $O_{uH}^{}$, $O_{dH}^{}$ and $O^{}_{\ell\ell}$ via the normalizations of the kinetic terms of $H$ and $\ell^{}_{\rm L}$ given in Eq.~(\ref{eq:shift}), the one-loop parts of EOMs in Eq. \eqref{eq:EOM-II}, and also the term in Eq. \eqref{eq:second-derivative}, i.e.,
\begin{eqnarray}
\delta C_{H}^{} &=& -3 C_{H}^{{\rm tree}} \delta Z_{H}^{} - 2 C^{\prime \rm G}_{HD}|^{}_{\rm tree} \left[ \delta \lambda^{\rm G} + \left( \lambda - 2\lambda^2_\Delta \right) \delta Z^{\rm G}_H \right] + \frac{12}{\left( 4\pi \right)^2 }\lambda^4_\Delta \left( 4\lambda^2_\Delta + \lambda^{}_3 \right) \left( 1 + L^{}_\Delta \right)
\nonumber
\\
&=&  \frac{- 4\lambda^2_\Delta}{\left(4\pi\right)^2} \left\{ \left( 3\lambda^2_3 + 2\lambda^2_4 \right) L^{}_\Delta + \left[ 16 \lambda \left( 8 + 5L^{}_\Delta \right) + 4 \left( 8\lambda^{}_1 + \lambda^{}_2 \right) \left( 1 + L^{}_\Delta \right) - \lambda^{}_3 \left( 52 + 19 L^{}_\Delta \right) \right.\right.
\nonumber
\\
&&  + \left.\left. \lambda^{}_4 (57 + 32L^{}_\Delta) \right]  \lambda^2_\Delta  - 4 \left( 67 + 23L^{}_\Delta \right) \lambda^4_\Delta \right\}\;,
\nonumber
\\
\delta C_{H\square}^{} &=& -2 C_{H\square}^{{\rm tree}} \delta Z_{H}^{}
= -\frac{24}{\left(4\pi\right)^2} \lambda_\Delta^4 \;,
\nonumber
\\
\delta C_{HD} &=& -2 \left(C_{HD}^{\rm tree}\right)^{\alpha\beta} \delta Z_{H} = -\frac{48}{(4\pi)^2} \lambda_\Delta^4 \;,
\nonumber
\\
\delta C_{eH}^{\alpha \beta }&=&-\frac{3}{2} \left(C_{eH}^{\rm tree}\right)^{\alpha\beta} \delta Z_{H} -\frac{1}{2} \left(\delta Z_{l}^\dag C_{eH}^{\rm tree}\right)^{\alpha\beta} - \frac{1}{2} C^{\prime \rm G}_{HD}|^{}_{\rm tree} \left( Y^{}_l \right)^{}_{\alpha\beta} \delta Z^{\rm G}_H 
\nonumber
\\
&=& -\frac{30}{\left(4\pi \right)^2} \lambda_\Delta^4\left(Y_l\right)_{\alpha\beta} - \frac{3}{4\left(4\pi \right)^2} (1+2L_\Delta) \lambda_\Delta^2\left( Y^{}_\Delta Y^\dag_\Delta Y^{}_l \right)_{\alpha \beta} \;,
\nonumber
\\
\delta C_{uH}^{\alpha \beta }&=&-\frac{3}{2} \left(C_{uH}^{\rm tree}\right)^{\alpha\beta} \delta Z_{H} - \frac{1}{2} C^{\prime \rm G}_{HD}|^{}_{\rm tree} \left( Y^{}_{\rm u} \right)^{}_{\alpha\beta} \delta Z^{\rm G}_H = -\frac{30}{(4\pi)^2 } \lambda_\Delta^4 \left( Y^{}_{\rm u} \right)_{\alpha\beta} \;,
\nonumber
\\
\delta C_{dH}^{\alpha \beta }&=&-\frac{3}{2} \left(C_{dH}^{\rm tree}\right)^{\alpha\beta} \delta Z_{H} - \frac{1}{2} C^{\prime \rm G}_{HD}|^{}_{\rm tree} \left( Y^{}_{\rm d} \right)^{}_{\alpha\beta} \delta Z^{\rm G}_H  = -\frac{30}{(4\pi)^2} \lambda_\Delta^4 \left( Y^{}_{\rm d} \right)_{\alpha\beta} \;,
\nonumber
\\
\delta C_{\ell \ell}^{\alpha\beta\gamma\delta} &=& - \frac{1}{8} \left[ \left( \delta Z_{l}^{\dag} Y^{}_\Delta \right)_{\alpha\gamma} \left( Y_\Delta^\dag \right)_{\beta\delta}
+ \left( \delta Z_{l}^\dag Y^{}_\Delta  \right)_{\gamma\alpha} \left( Y_\Delta^\dag \right)_{\beta\delta} + \left( Y^{}_\Delta \right)_{\alpha\gamma} \left( Y_\Delta^\dag \delta Z_{l}^{} \right)_{\delta\beta} \right.
\nonumber
\\
&& + \left. \left(  Y^{}_\Delta  \right)_{\alpha\gamma} \left( Y_\Delta^\dag\delta Z_{l} \right)_{\beta\delta} \right]
\nonumber
\\
&=& -\frac{3}{16(4\pi)^2} \left(1+2L_\Delta\right) \left[ \left( Y^{}_\Delta Y^\dag_\Delta Y^{}_\Delta \right)_{\alpha\gamma} \left(Y_\Delta^\dag\right)_{\beta\delta}  + \left( Y^{}_\Delta \right)_{\alpha\gamma} \left( Y_\Delta^\dag Y^{}_\Delta Y^\dag_\Delta \right)_{\beta\delta} \right] \;.
\label{eq:tree-to-one-loop}
\end{eqnarray} 
which will be added into the total one-loop-level Wilson coefficients of the corresponding operators.

All dim-6 operators in the Warsaw basis induced by integrating out the heavy triplet scalar at the one-loop level in the type-II seesaw model are listed in Table~\ref{tab:warsawbasis} and the associated one-loop-level Wilson coefficients up to $\mathcal{O} \left( M^{-2}_\Delta \right)$ are explicitly given in the remaining part of this subsection, where an overall loop factor $1/\left( 4\pi\right)^2$ is implied in all Wilson coefficients, and the contributions shown in Eq.~(\ref{eq:tree-to-one-loop}) have been added into the corresponding Wilson coefficients of $O^{}_H$, $O^{}_{H\square}$, $O_{HD}^{}$, $O_{eH}^{}$, $O_{uH}^{}$, $O_{dH}^{}$ and $O^{}_{\ell \ell}$.

\begin{table}
\hspace{0.05cm}
   \renewcommand\arraystretch{1.6}
      \scalebox{0.7}{
        \begin{tabular}{||c|c||c|c||c|c||}
            \hline \hline \multicolumn{2}{||c||}{$X^{3}$} & \multicolumn{2}{c||}{$H^{6} \text { and }
        H^{4} D^{2}$} & \multicolumn{2}{c||}{$\psi^{2} H^{3}$} \\
        \hline $O_{G}$ & $f^{A B C} G_{\mu}^{A \nu} G_{\nu}^{B \rho} G_{\rho}^{C \mu} $
        & \cellcolor{gray!30}{$O_{H}$} & \cellcolor{gray!30}{$\left(H^{\dagger} H\right)^{3}$} 
        & \cellcolor{gray!30}{$O_{e H}^{\alpha \beta }$} 
        & \cellcolor{gray!30}{$\left(H^{\dagger} H\right)\left(\overline{{\ell}^{}_{\alpha \text{L}}} E_{\beta\text{R}} H\right)$} \\
        $O_{\widetilde{G}}$ & $f^{A B C} \widetilde{G}_{\mu}^{A \nu} G_{\nu}^{B \rho} 
        G_{\rho}^{C \mu}$ & \cellcolor{gray!30}{$O_{H \square}$} & \cellcolor{gray!30}{$\left(H^{\dagger} H\right) 
        \square\left(H^{\dagger} H\right)$} & \cellcolor{gray!30}{$O_{u H}^{\alpha \beta }$} & \cellcolor{gray!30}
        {$\left(H^{\dagger} H\right)\left(\overline{Q_{\alpha \text{L}}} {U}^{}_{\beta\text{R}} \widetilde{H}\right)$} \\
        \cellcolor{gray!65}{$O_{W}$} & \cellcolor{gray!65}{$\color{blue}{\epsilon^{I J K} W_{\mu}^{I \nu} W_{\nu}^{J \rho} W_{\rho}^{K \mu}}$}
         & \cellcolor{gray!30}{$O_{H D}$}
         & \cellcolor{gray!30}{$\left(H^{\dagger} D^{\mu} H\right)^{\star}\left(H^{\dagger} D_{\mu} 
         H\right)$} & \cellcolor{gray!30}{$O_{d H}^{\alpha \beta } $}& \cellcolor{gray!30}{$\left(H^{\dagger} H\right)
         \left(\overline{Q_{\alpha \text{L}}} D_{\beta\text{R}} H\right)$}\\
        $O_{\widetilde{W}}$ & $\epsilon^{I J K} \widetilde{W}_{\mu}^{I \nu} 
        W_{\nu}^{J \rho} W_{\rho}^{K \mu}$ & & & & \\
        \hline \hline \multicolumn{2}{||c||}{$X^{2} H^{2}$} & \multicolumn{2}{c||}{$\psi^{2} X H$} & \multicolumn{2}
        {|c||}{$\psi^{2} H^{2} D$} \\
        \hline $O_{H G} $&$ H^{\dagger} H G_{\mu \nu}^{A} G^{A \mu \nu}$
         & \cellcolor{gray!30}{$O_{e W}^{\alpha \beta }$} &\cellcolor{gray!30}{ $\left(\overline{{\ell}^{}_{\alpha \text{L}}} \sigma^{\mu \nu} E_{\beta\text{R}}\right) \sigma^{I} H 
         W_{\mu \nu}^{I}$} & \cellcolor{gray!30}{$O_{H l}^{(1)\alpha\beta}$} & \cellcolor{gray!30}{$(H^{\dagger} \text{i} 
         \overleftrightarrow{D}_{\mu} H)\left(\overline{{\ell}^{}_{\alpha \text{L}}} \gamma^{\mu} 
         {\ell}^{}_{\beta\text{L}}\right)$} \\
       $ O_{H \widetilde{G}}$ & $H^{\dagger} H \widetilde{G}_{\mu \nu}^{A} 
        G^{A \mu \nu}$ & \cellcolor{gray!30}{$O_{e B}^{\alpha \beta }$} & \cellcolor{gray!30}{$\left(\overline{{\ell}^{}_{\alpha \text{L}}} \sigma^{\mu \nu} E_{\beta\text{R}}\right) H 
        B_{\mu \nu}$} & \cellcolor{gray!30}{$O_{H l}^{(3)\alpha \beta }$} & \cellcolor{gray!30}
        {$(H^{\dagger} \text{i} \overleftrightarrow{D}_{\mu}^IH)\left(\overline{{\ell}^{}_{\alpha \text{L}}} \sigma^{I} \gamma^{\mu} {\ell}^{}_{\beta\text{L}}\right)$} \\
        \cellcolor{gray!30}{$O_{H W}$}& \cellcolor{gray!30}{$H^{\dagger} H W_{\mu \nu}^{I} W^{I \mu \nu}$} 
        & $O_{u G}^{\alpha \beta }$ & $ \left(\overline{Q_{\alpha \text{L}}} \sigma^{\mu \nu} T^{A} {U}^{}_{\beta\text{R}}\right) \widetilde{H} G_{\mu \nu}^{A} $
        & \cellcolor{gray!30}{$O_{H e}^{\alpha \beta }$} & \cellcolor{gray!30}{$(H^{\dagger} \text{i} \overleftrightarrow{D}_{\mu} 
        H)\left(\overline{E_{\alpha\text{R}}} \gamma^{\mu} E_{\beta\text{R}}\right)$} \\
       $ O_{H \widetilde{W}}$ & $H^{\dagger} H \widetilde{W}_{\mu \nu}^{I} 
        W^{I \mu \nu}$ & $O_{u W}^{\alpha \beta }$ & $\left(\overline{Q_{\alpha \text{L}} }\sigma^{\mu \nu} {U}^{}_{\beta\text{R}}\right) \sigma^{I} 
        \widetilde{H} W_{\mu \nu}^{I}$ & \cellcolor{gray!30}{$O_{H q}^{(1)\alpha \beta }$} & \cellcolor{gray!30}{$(H^{\dagger} 
        \text{i} \overleftrightarrow{D}_{\mu} H)\left(\overline{Q_{\alpha \text{L}}} \gamma^{\mu} {Q}^{}_{\beta\text{L}}\right)$} \\
        \cellcolor{gray!30}{$O_{H B}$} & \cellcolor{gray!30}{$H^{\dagger} H B_{\mu \nu} B^{\mu \nu}$} & $O_{u B}^{\alpha \beta } $
        &$ \left(\overline{Q_{\alpha \text{L}}} \sigma^{\mu \nu} {U}^{}_{\beta \text{R}}\right) \widetilde{H} B_{\mu \nu} $
        & \cellcolor{gray!30}{$O_{H q}^{(3)\alpha \beta }$ }& \cellcolor{gray!30}{$(H^{\dagger} \text{i} \overleftrightarrow{D}_{\mu}^{I}
        H)\left(\overline{Q_{\alpha \text{L}}} \sigma^{I} \gamma^{\mu} {Q}^{}_{\beta\text{L}}\right) $}\\
        $O_{H \widetilde{B}}$ & $H^{\dagger} H \widetilde{B}_{\mu \nu} 
        B^{\mu \nu}$ & $O_{d G}^{\alpha \beta }$& $\left(\overline{Q_{\alpha \text{L}}} \sigma^{\mu \nu} T^{A} D_{\beta\text{R}}\right) 
        H G_{\mu \nu}^{A}$ & \cellcolor{gray!30}{$O_{H u}^{\alpha \beta }$} & \cellcolor{gray!30}
        {$(H^{\dagger} \text{i} \overleftrightarrow{D}_{\mu} H)\left(\overline{{U}^{}_{\alpha \text{R}}} \gamma^{\mu} {U}^{}_{\beta \text{R}}\right)$} \\
        \cellcolor{gray!30}{$O_{H W B}$} & \cellcolor{gray!30}{$H^{\dagger} \sigma^{I} H W_{\mu \nu}^{I} B^{\mu \nu}$}
         & $O_{d W}^{\alpha \beta } $
        & $\left(\overline{Q_{\alpha \text{L}}} \sigma^{\mu \nu} D_{\beta\text{R}}\right) \sigma^{I} H W_{\mu \nu}^{I}$ & \cellcolor{gray!30}{$O_{H d}^{\alpha \beta }$} 
        & \cellcolor{gray!30}{$(H^{\dagger} \text{i} \overleftrightarrow{D}_{\mu} H)\left(\overline{D_{\alpha \text{R}}} 
        \gamma^{\mu} D_{\beta\text{R}}\right)$} \\
       $ O_{H \widetilde{W} B} $& $H^{\dagger} \sigma^{I} H \widetilde{W}_{\mu \nu}^{I} 
        B^{\mu \nu} $& $O_{d B}^{\alpha \beta } $& $\left(\overline{Q_{\alpha \text{L}}} \sigma^{\mu \nu} D_{\beta\text{R}}\right) H B_{\mu \nu}$
         & $O_{H u d}^{\alpha \beta }$ & $\text{i}(\widetilde{H}^{\dagger} D_{\mu} 
         H)\left(\overline{{U}^{}_{\alpha \text{R}}} \gamma^{\mu} D_{\beta\text{R}}\right)$ \\
        \hline \hline \multicolumn{2}{||c||}
        {$\overline{{\text{L}}}\text{L}\overline{{\text{L}}^{}}\text{L}$} & \multicolumn{2}{c||} {$\overline{\text{R}}\text{R}\overline{\text{R}}\text{R}$} & \multicolumn{2}{c||}{$\overline{{\text{L}}^{}}\text{L}\overline{\text{R}}\text{R}$} \\
         \hline \rowcolor{gray!30} $O_{\ell\ell}^{\alpha \beta \gamma \delta} $& $\left({\overline{{\ell}^{}_{\alpha \text{L}}}}
         \gamma^\mu {\ell}^{}_{\beta\text{L}}\right)\left({\overline{{\ell}^{}_{\gamma \text{L}}}}\gamma^\mu {\ell}^{}_{\delta\text{L}}\right)$ 
        &\cellcolor{gray!65} $O_{ee}^{\alpha \beta \gamma \delta}$ &\cellcolor{gray!65} $\color{blue}{\left({\overline{E_{\alpha\text{R}}}}\gamma^\mu E_{\beta\text{R}}\right)
        \left({\overline{E_{\gamma\text{R}}}}\gamma^\mu E_{\delta\text{R}}\right) }$&$ O_{\ell e}^{\alpha \beta \gamma \delta}$
        &$ \left({\overline{{\ell}^{}_{\alpha \text{L}}}}\gamma^\mu {\ell}^{}_{\beta\text{L}}\right)
        \left({\overline{E_{\gamma\text{R}}}}\gamma^\mu E_{\delta\text{R}}\right)$ \\
        \cellcolor{gray!65} {$O_{qq}^{(1) \alpha \beta \gamma \delta} $}&\cellcolor{gray!65} $\color{blue}{\left({\overline{Q_{\alpha \text{L}}}}
        \gamma^\mu {Q}^{}_{\beta \text{L}}\right)\left({\overline{Q_{\gamma \text{L}}}}\gamma^\mu {Q}^{}_{\delta\text{L}}\right)}$
        &\cellcolor{gray!65} $O_{uu}^{\alpha \beta \gamma \delta} $&\cellcolor{gray!65} $\color{blue}{\left({\overline{{U}^{}_{\alpha \text{R}}}}\gamma^\mu 
        {U}^{}_{\beta\text{R}}\right)\left({\overline{{U}^{}_{\gamma \text{R}}}}\gamma^\mu {U}^{}_{\delta \text{R}}\right)}$ &\cellcolor{gray!30} $ O_{\ell u}^{\alpha \beta \gamma \delta} $
        &\cellcolor{gray!30} $\left({\overline{{\ell}^{}_{\alpha \text{L}}}}\gamma^\mu {\ell}^{}_{\beta\text{L}}\right)\left({\overline{{U}^{}_{\gamma \text{R}}}}\gamma^\mu {U}^{}_{\delta\text{R}}\right)$\\
        \rowcolor{gray!30} \cellcolor{gray!65} $O_{qq}^{(3)\alpha \beta \gamma \delta}$ & \cellcolor{gray!65} $\color{blue}{\left({\overline{Q_{\alpha \text{L}}}}
        \sigma^I\gamma^\mu {Q}^{}_{\beta\text{L}}\right)\left({\overline{Q_{\gamma \text{L}}}}
        \sigma^I\gamma^\mu {Q}^{}_{\delta\text{L}}\right) }$&\cellcolor{gray!65}$ O_{dd}^{\alpha \beta \gamma \delta}$
        &\cellcolor{gray!65} $\color{blue}{\left({\overline{D_{\alpha\text{R}}}}\gamma^\mu D_{\beta\text{R}}\right)
        \left({\overline{D_{\gamma\text{R}}}}\gamma^\mu D_{\delta\text{R}}\right)}$ & $O_{\ell d}^{\alpha \beta \gamma \delta}$ 
        & $\left({\overline{{\ell}^{}_{\alpha \text{L}}}}\gamma^\mu {\ell}^{}_{\beta\text{L}}\right)
        \left({\overline{D_{\gamma\text{R}}}}\gamma^\mu D_{\delta\text{R}}\right) $\\
        \rowcolor{gray!30} $O_{\ell q}^{(1)\alpha \beta \gamma \delta} $& $\left({\overline{{\ell}^{}_{\alpha \text{L}}}}
        \gamma^\mu {\ell}^{}_{\beta\text{L}}\right)\left({\overline{Q_{\gamma \text{L}}}}\gamma^\mu {Q}^{}_{\delta\text{L}}\right)$ & \cellcolor{gray!65}  $O_{eu}^{\alpha \beta \gamma \delta}$ 
        &  \cellcolor{gray!65} $\color{blue}{\left({\overline{E_{\alpha\text{R}}}}\gamma^\mu E_{\beta\text{R}}\right)\left({\overline{{U}^{}_{\gamma \text{R}}}}\gamma^\mu {U}^{}_{\delta\text{R}}\right) }$& \cellcolor{gray!65}  $O_{qe}^{\alpha \beta \gamma \delta}$
        & \cellcolor{gray!65} $\color{blue}{\left({\overline{Q_{\alpha \text{L}}}}\gamma^\mu {Q}^{}_{\beta\text{L}}\right)\left({\overline{E_{\gamma\text{R}}}}\gamma^\mu E_{\delta\text{R}}\right)}$ \\
        \rowcolor{gray!30} $O_{\ell q}^{(3)\alpha \beta \gamma \delta}$ &$\left({\overline{{\ell}^{}_{\alpha \text{L}}}}\sigma^I\gamma^\mu {\ell}^{}_{\beta\text{L}}\right)\left({\overline{Q_{\gamma \text{L}}}}\sigma^I\gamma^\mu {Q}^{}_{\delta\text{L}}\right)$
        & \cellcolor{gray!65} $O_{ed}^{\alpha \beta \gamma \delta}$& \cellcolor{gray!65}$\color{blue}{\left({\overline{E_{\alpha\text{R}}}}\gamma^\mu E_{\beta\text{R}}\right)\left({\overline{D_{\gamma\text{R}}}}\gamma^\mu D_{\delta\text{R}}\right)}$& $O_{qu}^{(1)\alpha \beta \gamma \delta}$
        &$\left({\overline{Q_{\alpha \text{L}}}}\gamma^\mu {Q}^{}_{\beta\text{L}}\right)\left({\overline{{U}^{}_{\gamma \text{R}}}}\gamma^\mu {U}^{}_{\delta\text{R}}\right)$\\
        & &\cellcolor{gray!65} $O_{ud}^{(1)\alpha \beta \gamma \delta}$ &\cellcolor{gray!65}{$\color{blue}{\left({\overline{{U}^{}_{\alpha \text{R}}}}\gamma^\mu {U}^{}_{\beta\text{R}}\right)
        \left({\overline{D_{\gamma\text{R}}}}\gamma^\mu D_{\delta\text{R}}\right)}$}
        &\cellcolor{gray!30}{$O_{qu}^{(8)\alpha \beta \gamma \delta}$}& \cellcolor{gray!30}{$\left({\overline{Q_{\alpha \text{L}}}}\gamma^\mu T^A{Q}^{}_{\beta\text{L}}\right)
        \left({\overline{{U}^{}_{\gamma \text{R}}}}\gamma^\mu T^A{U}^{}_{\delta\text{R}}\right)$}\\
        & &$O_{ud}^{(8)\alpha \beta \gamma \delta}$&$\left({\overline{{U}^{}_{\alpha \text{R}}}}\gamma^\mu T^A{U}^{}_{\beta\text{R}}\right)
        \left({\overline{D_{\gamma\text{R}}}}\gamma^\mu T^AD_{\delta\text{R}}\right)$
        &\cellcolor{gray!30}{$O_{qd}^{(1)\alpha \beta \gamma \delta}$}
        &\cellcolor{gray!30}{$\left({\overline{Q_{\alpha \text{L}}}}\gamma^\mu {Q}^{}_{\beta\text{L}}\right)\left({\overline{D_{\gamma\text{R}}}}\gamma^\mu D_{\delta}\right)$}\\
        & & & &\cellcolor{gray!30}{$O_{qd}^{(8)\alpha \beta \gamma \delta}$}&\cellcolor{gray!30}{$ \left({\overline{Q_{\alpha \text{L}}}}\gamma^\mu T^A{Q}^{}_{\beta\text{L}}\right)
        \left({\overline{D_{\gamma\text{R}}}}\gamma^\mu T^AD_{\delta\text{R}}\right)$}\\
        \hline \hline \multicolumn{2}{||c||}
        {$\left(\overline{{\text{L}}^{}}\text{R}\right)\left(\overline{\text{R}} \text{L} \right) \text{and} \left(\overline{{\text{L}}^{}}\text{R}\right)\left(\overline{{\text{L}}^{}}\text{R}\right)$} 
        & \multicolumn{4}{c||} {$B\text{-violating}$}\\
        \hline \cellcolor{gray!30}{$O_{\ell edq}^{\alpha \beta \gamma \delta}$} & \cellcolor{gray!30}{$\left({\overline{{\ell}^{}_{\alpha \text{L}}}}^jE_{\beta\text{R}}\right)
        \left({\overline{D_{\gamma\text{R}}}}{Q}_{\delta\text{L}}^j\right)$}
        & $ O_{duq}^{\alpha \beta \gamma \delta }$ & \multicolumn{3}{c||} {$\epsilon^{abc} \epsilon_{j k}\left[\left(D_{\alpha\text{R}}^{a}
        \right)^{\text{T}} {\sf C} {U}_{\beta\text{R}}^{b}\right]\left[\left({Q}_{\gamma\text{L}}^{c j}\right)^{\text{T}} {\sf C} {\ell}_{\delta\text{L}}^{k}\right] $}  \\
        \cellcolor{gray!30}{$O_{quqd}^{(1)\alpha \beta \gamma \delta}$}& \cellcolor{gray!30}{$\left({\overline{Q_{\alpha \text{L}}}}^j{U}^{}_{\beta\text{R}}\right)\epsilon_{jk}
        \left({\overline{Q}}_{\gamma}^kD_{\delta\text{R}}\right)$}
        & $O_{qqu}^{\alpha \beta \gamma \delta }$ & \multicolumn{3}{c||} {$\epsilon^{abc} \epsilon_{j k}\left[\left({Q}_{\alpha\text{L}}^{a j}
        \right)^{\text{T}} {\sf C} {Q}_{\beta\text{L}}^{b k}\right]\left[\left({U}_{\gamma\text{R}}^{c}\right)^{\text{T}} {\sf C} E_{\delta \text{R}}\right] $}  \\
        $O_{quqd}^{(8)\alpha \beta \gamma \delta}$ & $\left({\overline{Q_{\alpha}}}^j T^A {U}^{}_{\beta\text{R}}\right)\epsilon_{jk}\left({\overline{Q}}_{\gamma}^k T^A D_{\delta\text{R}}\right)$  
        & $ O_{qdd}^{\alpha \beta \gamma \delta }$ & \multicolumn{3}{c||} {$\epsilon^{abc} \epsilon_{j n} \epsilon_{k m}
        \left[\left({Q}_{\alpha \text{L}}^{a j}\right)^{\text{T}} {\sf C} {Q}_{\beta\text{L}}^{b k}\right]\left[\left({Q}_{\gamma\text{L}}^{c m}\right)^{\text{T}} {\sf C} {\ell}_{\delta\text{L}}^{n}\right]$} \\
        \cellcolor{gray!30}{$O_{\ell equ}^{(1)\alpha \beta \gamma \delta}$} & \cellcolor{gray!30}{$\left({\overline{{\ell}^{}_{\alpha \text{L}}}}^jE_{\beta \text{R}}\right)
        \epsilon_{jk}\left({\overline{Q_{\gamma \text{L}}}}^k{U}^{}_{\delta\text{R}}\right)$}
         & $ O_{duu}^{\alpha \beta \gamma \delta}$ & \multicolumn{3}{c||} {$\epsilon^{abc}\left[\left(D_{\alpha\text{R}}^{a}\right)^{\text{T}} 
        {\sf C} {U}_{\beta\text{R}}^{b}\right]\left[\left({U}_{\gamma\text{R}}^{c}\right)^{\text{T}}{\sf C}E_{\delta\text{R}}\right]$}\\ 
        $O_{\ell equ}^{(3)\alpha \beta \gamma \delta} $& $ \left({\overline{{\ell}^{}_{\alpha \text{L}}}}^j \sigma_{\mu\nu} E_{\beta\text{R}}\right)\epsilon_{jk}
        \left({\overline{Q_{\gamma \text{L}}}}^k \sigma^{\mu\nu}{U}^{}_{\delta\text{R}}\right)$ &   
        & \multicolumn{3}{c||} { } \\
        \hline \hline
    \end{tabular}
  }
\vspace{0.4cm}
\caption{Summary of the dimension-six operators in the Warsaw basis in the SMEFT~\cite{Grzadkowski:2010es}, where all the 41 operators induced at the tree and one-loop level by the heavy Higgs triplet in the type-II seesaw model are shown in the light gray region and those absent in the type-I seesaw model are further highlighted in blue in the dark gray region. The remaining 31 dimension-six operators are exactly those present in the low-energy EFT of the type-I seesaw model~\cite{Zhang:2021jdf}.}
\label{tab:warsawbasis}
\end{table}

\begin{itemize}
\item $X^3$
\begin{eqnarray}
C^{}_{W} &=& \frac{ g_2^3}{90} \;.
\label{eq:3W}
\end{eqnarray}
\item $X^2H^2$
\begin{eqnarray}
C^{}_{HB} &=& \frac{g_1^2}{4}\lambda_3-g_1^2\lambda^2_\Delta \;,
\label{eq:HBw}
\\
C^{}_{HWB} &=&  -\frac{g_1g_2}{3}\lambda_4-\frac{5g_1g_2}{3}\lambda^2_\Delta \;, 
\label{eq:HWBw}
\\
C^{}_{HW} &=&  \frac{g_2^2\ }{6}\lambda_3-\frac{g_2^2}{3}\lambda^2_\Delta \;.
\label{eq:HWw}
\end{eqnarray}

\item $H^4D^2$
\begin{eqnarray}
C^{}_{H\Box} &=& - \frac{g_1^4}{80} - \frac{g_2^4}{40} - \frac{\lambda_3^2}{4} + \frac{\lambda_4^2}{6} + \left[ \frac{g_1^2}{3} \left( 1 - 3L_\Delta \right) - \frac{g_2^2}{2} \left(9 + 14L_\Delta \right) + 8 \lambda \left( 3 + 2L_\Delta \right) \right.
\nonumber
\\
&& \left. + 4 \left( 8\lambda^{}_1 + \lambda^{}_2 \right) \left( 1+ L^{}_\Delta \right) - \lambda^{}_3 \left( 7 + 8L^{}_\Delta \right) + \frac{2}{3} \lambda_4 \left( 25 + 24L^{}_\Delta \right) \right] \lambda_{\Delta}^2
\nonumber
\\
&& - \frac{4}{3} \left( 67 + 24L_\Delta \right) \lambda_{\Delta}^4 \;,\qquad 
\label{eq:HBoxw}
\\
C^{}_{HD} &=& -\frac{g_1^4}{20}-\frac{2\lambda_4^2}{3}+\left[ \frac{g_1^2}{6} \left( 23 - 6L_\Delta \right) +\frac{g_2^2}{2} \left( 5 + 6L_\Delta \right) + 4 \left( 3 + 2L_\Delta \right) \lambda \right.
\nonumber
\\
&& \left. + 8 \left( 8\lambda^{}_1 + \lambda^{}_2 - 2\lambda^{}_3 \right) \left( 1+ L^{}_\Delta\right) -\frac{8}{3} \lambda_4 \left( 11 + 12L^{}_\Delta  \right) \right] \lambda_{\Delta}^2 -\frac{16}{3}(17+3L_\Delta)\lambda_\mathrm{\Delta}^4 \;. 
\label{eq:HDw}
\end{eqnarray}

\item $H^6$
\begin{eqnarray}
C^{}_H &=& -\frac{g_2^4\lambda}{15}-\frac{\lambda_3^3}{2}+\frac{4\lambda\lambda_4^2}{3}-\lambda_3\lambda_4^2+\left[ \frac{2g_2^4}{15} - g_1^2 \lambda \left( 5 + 6 L_\Delta \right) -\frac{1}{3} g_2^2 \lambda \left(61+86L_\Delta\right)
\right.
\nonumber
\\
&&  + 120\lambda^2 \left( 3 + 2L_\Delta \right) + 16 \lambda \left( 8 \lambda^{}_1 + \lambda^{}_2 \right) \left( 1+ L^{}_\Delta \right)- 60 \lambda\lambda_3 \left(3+2L_\Delta\right) 
\nonumber
\\
&& +\frac{8}{3} \lambda\lambda_4 \left( 83 + 60 L^{}_\Delta \right) - 2 \lambda_3 \left( 8\lambda^{}_1 + \lambda^{}_2 + 16 \lambda^{}_4\right) \left( 2 + L^{}_\Delta \right) + 4 \lambda_1\lambda_4 \left( 8 + 7L^{}_\Delta \right) 
\nonumber
\\
&&  + \left. 2 \lambda_2\lambda_4 \left( 2 + 3L^{}_\Delta \right) + 26 \lambda_3^2 +\frac{4}{3} \lambda_4^2 \left( 25 + 12L^{}_\Delta \right) \right] \lambda_\Delta^2
+ \frac{2}{3} \lambda^4_\Delta \left[ 3g^2_1 \left( 5+6L^{}_\Delta \right) \right.
\nonumber
\\
&& + g^2_2 \left(61 + 86L^{}_\Delta \right) - 32\lambda \left( 107 + 60L^{}_\Delta \right) - 24 \lambda^{}_1 \left( 29 + 24L^{}_\Delta \right) - 72\lambda^{}_2 \left( 1+L^{}_\Delta \right) 
\nonumber
\\
&& + \left. 6\lambda^{}_3 \left( 171 + 82L^{}_\Delta \right) - 2\lambda^{}_4 \left(605 + 342L^{}_\Delta \right) \right] + \frac{128}{3}\lambda^6_\Delta \left( 71 + 30 L^{}_\Delta \right) \nonumber \\
&& + 12 \lambda^4_\Delta \left( 4\lambda^2_\Delta + \lambda^{}_3 \right) \left( 1+ L^{}_\Delta \right) \;.
\label{eq:H6}
\end{eqnarray}

\item $\psi^2 X H$
\begin{eqnarray}
C^{\alpha\beta}_{eB} &=&  +\frac{g_1}{4}\left(Y_\Delta^{{}}Y_\Delta^\dag Y_l^{{}}\right)_{\alpha\beta} \;,
\label{eq:eBw}
%\end{eqnarray}
%\begin{eqnarray}
\\
C^{\alpha\beta}_{eW} &=& -\frac{g_2}{8}\left(Y_\Delta^{{}}Y_\Delta^\dag Y_l^{{}}\right)_{\alpha\beta} \;.
\label{eq:eWw}
\end{eqnarray}

\item $\psi^2 D H^2$
\begin{eqnarray}
C^{(1)\alpha\beta}_{Hq} &=& -\frac{g_1^4}{120}\delta_{\alpha\beta}+\frac{g_1^2}{36}\left(19+6L_\Delta\right)\lambda^2_\Delta\delta_{\alpha\beta}-\frac{3}{4}\left(5+2L_\Delta\right)\lambda^2_\Delta\left(Y_{\rm d}^{{}}Y_{\rm d}^\dag-Y_{\rm u}^{{}}Y_{\rm u}^\dag\right)_{\alpha\beta} \;,
\label{eq:HQ1w}
\\
C^{(3)\alpha\beta}_{Hq} &=& -\frac{g_2^4}{60}\delta_{\alpha\beta}+\frac{g_2^2}{12}\left(7+2L_\Delta\right)\lambda^2_\Delta\delta_{\alpha\beta}-\frac{1}{4}\left(5+2L_\Delta\right)\lambda^2_\Delta\left(Y_{\rm d}^{{}}Y_{\rm d}^\dag+Y_{\rm u}^{{}}Y_{\rm u}^\dag\right)_{\alpha\beta} \;,
\label{eq:HQ3w}
\\
C^{\alpha\beta}_{Hu} &=& -\frac{g_1^4}{30}\delta_{\alpha\beta}+\frac{g_1^2}{9}\left(19+6L_\Delta\right)\lambda^2_\Delta\delta_{\alpha\beta}-\frac{3}{2}\left(5+2L_\Delta\right)\lambda^2_\Delta\left(Y_{\rm u}^\dag Y_{\rm u}^{{}}\right)_{\alpha\beta} \;,
\label{eq:HUw}
\\
C^{\alpha\beta}_{Hd} &=& \frac{g_1^4}{60}\delta_{\alpha\beta}-\frac{g_1^2}{18}\left(19+6L_\Delta\right)\lambda^2_\Delta\delta_{\alpha\beta}+\frac{3}{2}\left(5+2L_\Delta\right)\lambda^2_\Delta\left(Y_{\rm d}^\dag Y_{\rm d}^{{}}\right)_{\alpha\beta} \;,
\label{eq:HDDw}
\end{eqnarray}
\vspace{-0.8cm}
\begin{eqnarray}
C^{(1)\alpha\beta}_{H\ell} &=& \frac{g_1^4}{40}\delta_{\alpha\beta}-\frac{g_1^2}{12}\left(19+6L_\Delta\right)\lambda^2_\Delta\delta_{\alpha\beta}-\frac{g_1^2}{12}\left(5+3L_\Delta\right)\left(Y_\Delta^{{}}Y_\Delta^\dag\right)_{\alpha\beta}
\nonumber
\\
&&-\frac{3}{4}\left[ \left(5+2L_\Delta\right) \lambda^2_\Delta \left( Y_l^{{}}Y_l^\dag + 2 Y_\Delta^{} Y_\Delta^\dag \right)_{\alpha\beta} + \left( 1 + L_\Delta \right) \left( Y_\Delta^{} Y_l^{\ast} Y_l^{\rm T} Y_\Delta^{\dagger} \right)_{\alpha\beta} \right] \;,
\label{eq:Hl1w}
%\end{eqnarray}
%\vspace{-0.8cm}
%\begin{eqnarray}
\\
C^{(3)\alpha\beta}_{H\ell} &=& -\frac{g_2^4}{60}\delta_{\alpha\beta}+\frac{g_2^2}{12}\left(7+2L_\Delta\right)\lambda^2_\Delta\delta_{\alpha\beta}+\frac{g_2^2}{12}\left(2+L_\Delta\right)\left(Y_\Delta^{{}}Y_\Delta^\dag\right)_{\alpha\beta}
\nonumber
\\
&& -\frac{1}{4}\left[ \left(5+2L_\Delta\right)\lambda^2_\Delta \left( Y_l^{{}}Y_l^\dag - 4 Y^{}_\Delta Y^\dagger_\Delta \right)_{\alpha\beta} + \left(1+L_\Delta\right)\left( Y_\Delta^{} Y_l^{\ast} Y_l^{\rm T} Y_\Delta^{\dagger} \right)_{\alpha\beta}\right] \;,
\label{eq:Hl3w}
\\
C^{\alpha\beta}_{He} &=& \frac{g_1^4}{20}\delta_{\alpha\beta}-\frac{g_1^2}{6}\left(19+6L_\Delta\right)\lambda^2_\Delta\delta_{\alpha\beta} + \frac{3}{2} \left(5+2L_\Delta\right) \lambda^2_\Delta \left(Y_l^\dag Y_l^{{}}\right)_{\alpha\beta} 
\nonumber
\\
&& +\frac{1}{4}\left(Y_l^\dag Y_\Delta^{{}}Y_\Delta^\dag Y_l^{{}}\right)_{\alpha\beta} \;.
\label{eq:Hew}
\end{eqnarray}

\item $\psi^2H^3$
\begin{eqnarray}
C^{\alpha\beta}_{uH} &=& \left\{  -\frac{g_2^4}{60} + \frac{\lambda_4^2}{3}+\left[-\frac{1}{4}g_1^2\left(5+6L_\Delta\right)-\frac{1}{12}g_2^2\left(61+86L_\Delta\right) +16\lambda (3 + L^{}_\Delta) \right.\right.
\nonumber
\\
&& \left.  + 4\left( 8\lambda^{}_1 + \lambda^{}_2 \right) \left( 1+ L^{}_\Delta \right) - \lambda^{}_3 \left( 13 + 8L^{}_\Delta \right) + \frac{2}{3} \lambda^{}_4 \left( 35 + 24L^{}_\Delta \right) \right] \lambda_{\Delta}^2 
\nonumber
\\
&& - \left. \frac{2}{3} \lambda_{\Delta}^4 \left(205 + 48 L_\Delta \right) \right\}  \left(Y_{\rm u}\right)_{\alpha\beta} - \frac{1}{2} \lambda^2_\Delta \left[ \left( 7 + 6L^{}_\Delta \right) \left( Y_{\rm d}^{{}} Y_{\rm d}^\dag Y_{\rm u}^{}\right)_{\alpha\beta} \right.
\nonumber
\\
&&- \left. 5 \left(1+2L_\Delta\right)  \left(Y_{\rm u}^{{}} Y_{\rm u}^\dag Y_{\rm u}^{{}} \right)_{\alpha\beta} \right] \;, \qquad
\label{eq:UHw}
\\
C^{\alpha\beta}_{dH} &=& \left\{  -\frac{g_2^4}{60} + \frac{\lambda_4^2}{3}+\left[-\frac{1}{4}g_1^2\left(5+6L_\Delta\right)-\frac{1}{12}g_2^2\left(61+86L_\Delta\right) +16\lambda ( 3 + L^{}_\Delta) \right.\right.
\nonumber
\\
&& \left.  + 4\left( 8\lambda^{}_1 + \lambda^{}_2 \right) \left( 1+ L^{}_\Delta \right) - \lambda^{}_3 \left( 13 + 8L^{}_\Delta \right) + \frac{2}{3} \lambda^{}_4 \left( 35 + 24L^{}_\Delta \right) \right] \lambda_{\Delta}^2 
\nonumber
\\
&& - \left. \frac{2}{3} \lambda_{\Delta}^4 \left(205 + 48 L_\Delta \right) \right\}  \left(Y_{\rm d}\right)_{\alpha\beta} - \frac{1}{2} \lambda^2_\Delta \left[ \left( 7 + 6L^{}_\Delta \right) \left( Y_{\rm u}^{{}} Y_{\rm u}^\dag Y_{\rm d}^{}\right)_{\alpha\beta} \right.
\nonumber
\\
&&- \left. 5 \left(1+2L_\Delta\right)  \left(Y_{\rm d}^{{}} Y_{\rm d}^\dag Y_{\rm d}^{{}} \right)_{\alpha\beta} \right] \;, \qquad
\label{eq:DHw}
\\
C^{\alpha\beta}_{eH} &=& \left\{  -\frac{g_2^4}{60} + \frac{\lambda_4^2}{3}+\left[-\frac{1}{4}g_1^2\left(5+6L_\Delta\right)-\frac{1}{12}g_2^2\left(61+86L_\Delta\right) +16\lambda (3 + L^{}_\Delta) \right.\right.
\nonumber
\\
&& \left.  + 4\left( 8\lambda^{}_1 + \lambda^{}_2 \right) \left( 1+ L^{}_\Delta \right) - \lambda^{}_3 \left( 13 + 8L^{}_\Delta \right) + \frac{2}{3} \lambda^{}_4 \left( 35 + 24L^{}_\Delta \right) \right] \lambda_{\Delta}^2 
\nonumber
\\
&& - \left. \frac{2}{3} \lambda_{\Delta}^4 \left(205 + 48 L_\Delta \right) \right\}  \left(Y_l \right)_{\alpha\beta} + \frac{5}{2} \lambda^2_\Delta  \left(1+2L_\Delta\right)  \left(Y_l^{{}} Y_l^\dag Y_l^{{}} \right)_{\alpha\beta}  
\nonumber
\\
&& - \frac{1}{4} \left[ 3\lambda^{}_3 + 2\lambda^{}_4 - \lambda^2_\Delta \left( 7+6L^{}_\Delta \right) \right] \left( Y^{}_\Delta Y^\dagger_\Delta Y^{}_l \right)^{}_{\alpha\beta} - \frac{1}{2} \left( Y^{}_\Delta Y^\ast_l Y^{\rm T} Y^\dagger_\Delta Y^{}_l \right)^{}_{\alpha\beta}
\nonumber
\\
&& + \frac{1}{4} \left( Y^{}_l Y^\dagger_l Y^{}_\Delta Y^\dagger_\Delta Y^{}_l \right)^{}_{\alpha\beta} \;.\qquad\quad
\label{eq:eHw}
\end{eqnarray}

\item Four-quark
\begin{eqnarray}
C^{(1)\alpha\beta\gamma\delta}_{qq} &=&-\frac{g_1^4}{720}\delta_{\alpha\beta}\delta_{\gamma\delta} \;,
\label{eq:qq1}
\\
C^{(3)\alpha\beta\gamma\delta}_{qq} &=&-\frac{g_2^4}{120}\delta_{\alpha\beta}\delta_{\gamma\delta} \;,
\label{eq:qq3}
\\
C^{\alpha\beta\gamma\delta}_{uu} &=&-\frac{g_1^4}{45}\delta_{\alpha\beta}\delta_{\gamma\delta} \;,
\label{eq:uu}
\\
C^{\alpha\beta\gamma\delta}_{dd} &=&-\frac{g_1^4}{180}\delta_{\alpha\beta}\delta_{\gamma\delta} \;,
\label{eq:dd}
\\
C^{(1)\alpha\beta\gamma\delta}_{ud} &=&\frac{g_1^4}{45}\delta_{\alpha\beta}\delta_{\gamma\delta} \;,
\label{eq:ud}
\\
C^{(1)\alpha\beta\gamma\delta}_{qu} &=&-\frac{g_1^4}{90}\delta_{\alpha\beta}\delta_{\gamma\delta}-\frac{1}{3}\lambda^2_\Delta\left(Y_{\rm u}^{{}}\right)_{\alpha\delta}\left(Y_{\rm u}^\dag\right)_{\gamma\beta} \;,
\label{eq:QU1w}
\\
C^{(8)\alpha\beta\gamma\delta}_{qu} &=& -2\left(Y_{\rm u}^{{}}\right)_{\alpha\delta}\left(Y_{\rm u}^\dag\right)_{\gamma\beta}\lambda^2_\Delta \;,
\label{eq:QU8w}
\\
C^{(1)\alpha\beta\gamma\delta}_{qd} &=& \frac{g_1^4}{180}\delta_{\alpha\beta}\delta_{\gamma\delta}-\frac{1}{3}\lambda^2_\Delta\left(Y_{\rm d}^{{}}\right)_{\alpha\delta}\left(Y_{\rm d}^\dag\right)_{\gamma\beta} \;,
\label{eq:Qd1w}
\\
C^{(8)\alpha\beta\gamma\delta}_{qd} &=& -2\left(Y_{\rm d}^{{}}\right)_{\alpha\delta}\left(Y_{\rm d}^\dag\right)_{\gamma\beta}\lambda^2_\Delta \;,
\label{eq:Qd8w}
\\
C^{(1)\alpha\beta\gamma\delta}_{quqd} &=& 2\left(Y_{\rm u}^{{}}\right)_{\alpha\beta}\left(Y_{\rm d}^{{}}\right)_{\gamma\delta}\lambda^2_\Delta \;.
\label{eq:QUQdw}
\end{eqnarray}

\item Four-lepton
\begin{eqnarray}
C^{\alpha\beta\gamma\delta}_{\ell\ell} &=& - \frac{g_2^4}{120} \left( 2\delta_{\alpha\delta} \delta_{\gamma\beta} - \delta_{\alpha\beta} \delta_{\gamma\delta} \right) - \frac{g_1^4}{80} \delta_{\alpha\beta} \delta_{\gamma\delta}  + \frac{g_1^2}{12} \left(5+3 L_\Delta\right)(Y_\Delta^{{}}Y_\Delta^\dag)^{}_{\alpha\beta} \delta^{}_{\gamma\delta}
\nonumber
\\
&& + \frac{g^2_2}{12} \left(2+L_\Delta\right) \left[ 2 \left(Y_\Delta^{{}}Y_\Delta^\dag\right)^{}_{\alpha\beta} \delta^{}_{\gamma\delta} + 2 \left( Y^{}_\Delta Y^\dagger_\Delta \right)^{}_{\alpha\delta} \delta^{}_{\gamma\beta} \right] - \frac{1}{8} (Y_\Delta^{{}}Y_\Delta^\dag)^{}_{\alpha\beta} (Y_\Delta^{{}}Y_\Delta^\dag)_{\gamma\delta}
\nonumber
\\
&& - \frac{1}{2} (Y_\Delta^{{}}Y_\Delta^\dag)_{\alpha\delta} (Y_\Delta^{{}}Y_\Delta^\dag)_{\gamma\beta} + \frac{1}{4} \left(8\lambda_1+\lambda_2\right) \left(1+L_\Delta\right) (Y_\Delta^{})_{\alpha\gamma}(Y_\Delta^\dagger)_{\delta\beta}
\nonumber
\\
&& - \frac{3}{16} \left( 1+2L_\Delta \right)\left[ \left(Y^{}_\Delta Y^\dag_\Delta Y^{}_\Delta \right)_{\alpha\gamma} \left( Y_\Delta^\dag \right)_{\beta\delta} + \left( Y^{}_\Delta \right)_{\alpha\gamma} \left( Y_\Delta^\dag Y^{}_\Delta Y^\dag_\Delta \right)_{\beta\delta} \right] \;,
\label{eq:llw}
\\
C^{\alpha\beta\gamma\delta}_{\ell e} &=& 
-\frac{g_1^4}{20}\delta_{\alpha\beta}\delta_{\gamma\delta}-\lambda^2_\Delta\left(Y_l^{{}}\right)_{\alpha\delta}\left(Y_l^\dag\right)_{\gamma\beta}+\frac{g_1^2}{6}\left(5+3L_\Delta\right)\left(Y_\Delta^{{}}Y_\Delta^\dag\right)_{\alpha\beta}\delta_{\gamma\delta} 
\nonumber
\\
&&-\frac{3}{8}{\left(3+2L_\Delta\right)\left(Y_l^\dag Y_\Delta^{{}}\right)}_{\gamma \alpha}\left(Y_\Delta^\dag Y_l^{{}}\right)_{\beta\delta} \;
\label{eq:lew}
\\
C^{\alpha\beta\gamma\delta}_{e e} &=& -\frac{g_1^4}{20}\delta_{\alpha\beta}\delta_{\gamma\delta} \;.\qquad\;
\label{eq:eew}
\end{eqnarray}

\item Semileptonic
\begin{eqnarray}
C^{(1)\alpha\beta\gamma\delta}_{\ell q} &=&\frac{g_1^4}{120}\delta_{\alpha\beta}\delta_{\gamma\delta}-\frac{g_1^2}{36}\left(5+3L_\Delta\right)\left(Y_\Delta^{{}}Y_\Delta^\dag\right)_{\alpha\beta}\delta_{\gamma\delta} \;,
\label{eq:lQ1w}
\\
C^{(3)\alpha\beta\gamma\delta}_{\ell q} &=& -\frac{g_2^4}{60}\delta_{\alpha\beta}\delta_{\gamma\delta}+\frac{g_2^2}{12}\left(2+L_\Delta\right)\left(Y_\Delta^{{}}Y_\Delta^\dag\right)_{\alpha\beta}\delta_{\gamma\delta} \;,\qquad
\label{eq:lQ3w}
\\
C^{\alpha\beta\gamma\delta}_{eu} &=&\frac{g_1^4}{15}\delta_{\alpha\beta}\delta_{\gamma\delta}\;,
\label{eq:eu}
\\
C^{\alpha\beta\gamma\delta}_{ed} &=&-\frac{g_1^4}{30}\delta_{\alpha\beta}\delta_{\gamma\delta} \;,
\label{eq:ed}
\\
C^{\alpha\beta\gamma\delta}_{qe} &=&\frac{g_1^4}{60}\delta_{\alpha\beta}\delta_{\gamma\delta} \;,
\label{eq:qe}
\\
C^{\alpha\beta\gamma\delta}_{\ell u} &=& \frac{g_1^4}{30}\delta_{\alpha\beta}\delta_{\gamma\delta}-\frac{g_1^2}{9}\left(5+3L_\Delta\right)\left(Y_\Delta^{{}}Y_\Delta^\dag\right)_{\alpha\beta}\delta_{\gamma\delta} \;,
\label{eq:lUw}
\\
C^{\alpha\beta\gamma\delta}_{\ell d} &=& -\frac{g_1^4}{60}\delta_{\alpha\beta}\delta_{\gamma\delta}+\frac{g_1^2}{18}\left(5+3L_\Delta\right)\left(Y_\Delta^{{}}Y_\Delta^\dag\right)_{\alpha\beta}\delta_{\gamma\delta} \;,
\label{eq:ldw}
\\
C^{\alpha\beta\gamma\delta}_{\ell e d q} &=& 2\left(Y_l^{{}}\right)_{\alpha\beta}\left(Y_{\rm d}^\dag\right)_{\gamma\delta}\lambda^2_\Delta \;,
\label{eq:ledQw}
\\
C^{(1)\alpha\beta\gamma\delta}_{\ell e q u} &=& -2\left(Y_l^{{}}\right)_{\alpha\beta}\left(Y_{\rm u}^{{}}\right)_{\gamma\delta}\lambda^2_\Delta \;.
\label{eq:leQUw}
\end{eqnarray}

\end{itemize}

Now, with all above results, we can write out the complete Lagrangian of the SEFT-II up to the one-loop level, that is
\begin{eqnarray}
    \mathcal{L}^{}_{\rm SEFT-II} &=& \mathcal{L}^{}_{\rm SM} \left(m^2 \to m^2_{\rm eff}, \lambda\to \lambda_{\rm eff}, Y_l^{}\to Y_l^{\rm eff}, Y_{\rm u}^{}\to Y_{\rm u}^{\rm eff},Y_{\rm d}^{}\to Y_{\rm d}^{\rm eff}, g^{}_1 \to g^{\rm eff}_1, g^{}_2 \to g^{\rm eff}_2 \right) 
    \nonumber
    \\
    && +\frac{1}{M^{}_\Delta}  \left[ \left(C^{(5)}_{\rm eff}\right)_{\alpha\beta} O^{(5)}_{\alpha \beta} + {\rm h.c.} \right] + \frac{C^{\rm tree}_{H}}{M^2_\Delta} O^{}_{H} + \frac{C^{\rm tree}_{H\square}}{M^2_\Delta} O^{}_{H\square} + \frac{C^{\rm tree}_{HD}}{M^2_\Delta} O^{}_{HD}
    \nonumber
    \\
    && + \frac{1}{M^2_\Delta} \left[ \left(C^{\rm tree}_{eH}\right)^{}_{\alpha\beta} O^{\alpha\beta}_{eH}  + \left(C^{\rm tree}_{uH}\right)^{}_{\alpha\beta} O^{\alpha\beta}_{uH} + \left(C^{\rm tree}_{dH}\right)^{}_{\alpha\beta} O^{\alpha\beta}_{dH} + {\rm h.c.} \right] 
    \nonumber
    \\
     && + \frac{\left(C^{\rm tree}_{\ell\ell}\right)^{}_{\alpha\beta\gamma\delta}}{M^2_\Delta} O^{\alpha\beta\gamma\delta}_{\ell\ell} + \sum^{}_i \frac{C^{}_i}{M^2_\Delta} O^{}_i \;, \quad
    %\frac{\lambda^{}_\Delta \left( Y^{}_\Delta \right)^{}_{\alpha\beta}}{M^{}_\Delta } \overline{\ell^{}_{\alpha \rm L}} \widetilde{H} \widetilde{H}^T \ell^c_{\beta \rm L}
    \label{eq:lagrangian-eft}
\end{eqnarray}
where the couplings in the SM have been replaced by the effective coupling given in Eqs.~\eqref{eq:g-eff} and \eqref{eq:threshold-corrections}, $C^{(5)}_{\rm eff}$ is given in Eq.~\eqref{eq:WC-dim-5} and includes both the tree-level and one-loop contributions, $C^{\rm tree}_H, C^{\rm tree}_{H\square}, C^{\rm tree}_{HD}, C^{\rm tree}_{eH}, C^{\rm tree}_{uH}, C^{\rm tree}_{dH}$ and $C^{\rm tree}_{\ell \ell}$ are the tree-level contributions to the Wilson coefficients of the associated dim-6 operators and are listed in Eq.~\eqref{eq:tree-WC}, $O^{}_i$ denote the dim-6 operators listed in Table~\ref{tab:warsawbasis} including Hermitian conjugations of the non-Hermitian operators, and $C^{}_i$ refer to the one-loop contributions to the corresponding Wilson coefficients. Unlike what we have done for the unique dimension-five operator, the tree- and one-loop-level contributions to the Wilson coefficients of $O^{}_H$, $O^{}_{H\square}$, $O_{HD}^{}$, $O_{eH}^{}$, $O_{uH}^{}$, $O_{dH}^{}$ and $O^{}_{\ell \ell}$ are not summed up.

\subsection{Diagrammatic approach}

The one-loop matching of an UV theory onto the SMEFT can also be achieved by calculating the 1LPI Feynman diagrams and extracting the hard-part contributions of the amplitudes by virtue of expansion by regions. As an independent cross-check of the functional approach, the tree-level and one-loop matching of a general UV model onto arbitrary effective theories by diagrammatic approach has recently been automated in the publicly available package {\sf Matchmakereft}\cite{Carmona:2021xtq}. It is worthwhile to mention that 
the one-loop matching of the type-I seesaw model onto the SMEFT has been done in Ref.~\cite{Ohlsson:2022hfl} through the diagrammatic calculations, which should be compared with the functional approach in Ref.~\cite{Zhang:2021jdf}.

\begin{figure}[t]
\centering
\includegraphics[width=1\textwidth]{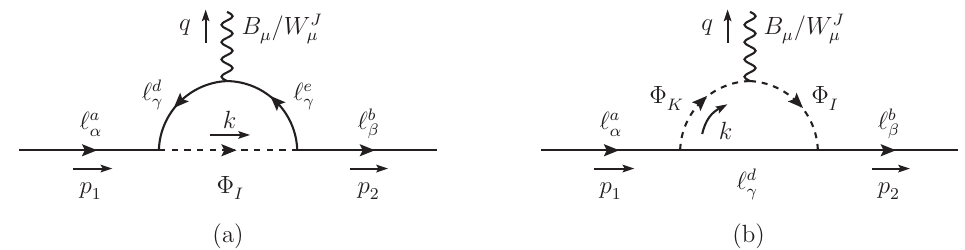}
\vspace{-0.3cm}
\caption{Feynman diagrams for the amplitude $\langle \ell\ell S \rangle$ in the UV model, which should be matched by operators $O^{}_{\ell D}$, $O^{}_{S\ell}$, $O^\prime_{S\ell}$ and $O^\prime_{\widetilde{S}\ell}$ (for $S=B,W$) in the Green's basis in the EFT.}
\label{fig:diagrams-matching}
\end{figure}

Since we will calculate the radiative decays of charged lepton $l^-_\alpha \to l^-_\beta + \gamma$ in the SEFT-II in the next section, it is instructive to make use of the Feynman diagrammatic approach to match the relevant operators, i.e., $O^{}_{eB}$ and $O^{}_{eW}$ in the Warsaw basis, following the procedure in Ref.~\cite{Zhang:2021tsq}. In the type-II seesaw model, only the operators, i.e.,
\begin{eqnarray}
&&O^{\alpha\beta}_{\ell D} = \frac{{\rm i}}{2} \overline{\ell^{}_{\alpha{\rm L}}} \left(D^2 \slashed{D} + \slashed{D} D^2 \right) \ell^{}_{\beta {\rm L}} \;,
\nonumber
\\
&&O^{\prime\alpha\beta}_{B\ell} = \frac{\rm i}{2} \overline{\ell^{}_{\alpha{\rm L}}} \gamma^\mu \Dlrn \ell^{}_{\beta {\rm L}} B^{}_{\mu\nu} \;,\quad O^{\prime\alpha\beta}_{W\ell} = \frac{\rm i}{2} \overline{\ell^{}_{\alpha{\rm L}}} \gamma^\mu \Dilrn \ell^{}_{\beta {\rm L}} W^I_{\mu\nu} \;,
\nonumber
\\
&&O^{\prime\alpha\beta}_{\widetilde{B}\ell} =\frac{\rm i}{2} \overline{\ell^{}_{\alpha{\rm L}}} \gamma^\mu \Dlrn \ell^{}_{\beta {\rm L}} \widetilde{B}^{}_{\mu\nu} \;,\quad O^{\prime\alpha\beta}_{\widetilde{W}\ell} =\frac{\rm i}{2} \overline{\ell^{}_{\alpha{\rm L}}} \gamma^\mu \Dilrn \ell^{}_{\beta {\rm L}} \widetilde{W}^I_{\mu\nu} \;,
\label{eq:matching-1}
\end{eqnarray}
in the Green's basis contribute to $O^{}_{eB}$ and $O^{}_{eW}$, whereas others in the Green's basis contributing to $O^{}_{eB}$ and $O^{}_{eW}$ are absent due to the special interactions of the triplet scalar. Thus we need only to derive the Wilson coefficients of those operators listed in Eq.~\eqref{eq:matching-1} via the Feynman diagrammatic approach. For this purpose, we calculate the amplitudes corresponding to the Feynman diagrams (a) and (b) in Fig.~\ref{fig:diagrams-matching}. 
There are another two operators in the Green's basis contributing to these amplitudes, namely,\footnote{The threshold correction to the kinetic term of the lepton doublet, i.e., $\delta Z^{\rm G}_\ell \overline{\ell^{}_{\rm L}} {\rm i} \slashed{D} \ell^{}_{\rm L}$ in the Green's basis, also contributes to those amplitudes from Fig.~\ref{fig:diagrams-matching}. However, since the contributions from this dim-4 term can be easily distinguished from those induced by the dim-6 operators given in Eqs.~\eqref{eq:matching-1} and \eqref{eq:matching-2}. Hence we do not consider this threshold correction here.}
\begin{eqnarray}
O^{\alpha\beta}_{B\ell} = \overline{\ell^{}_{\alpha{\rm L}}} \gamma^\mu \ell^{}_{\beta {\rm L}} D^\nu B^{}_{\mu\nu} \;,\quad O^{\alpha\beta}_{W\ell} =  \overline{\ell^{}_{\alpha{\rm L}}} \gamma^\mu \sigma^I \ell^{}_{\beta {\rm L}} \left(D^\nu W^{}_{\mu\nu}\right)^I \;.
\label{eq:matching-2}
\end{eqnarray}
When the one-loop matching is carried out by calculating the Feynman diagrams in Fig.~\ref{fig:diagrams-matching}, those two operators in Eq.~\eqref{eq:matching-2} should also be taken into consideration, though they do not contribute to $O^{}_{eB}$ and $O^{}_{eW}$ in the Warsaw basis.  

%%%%%%%%%%%%%%%%%%%%%%%%%%figure 1 %%%%%%%%%%%%%%%%%%%%%%%%%%
%%%%%%%%%%%%%%%%%%%%%%%%%%%%%%%%%%%%%%%%%%%%%%%%%%%%%%%%%%%%%

The contributions from the operators given in Eqs.~\eqref{eq:matching-1} and \eqref{eq:matching-2} to the amplitudes $\langle \ell \ell B \rangle$ and $\langle \ell \ell W \rangle$ in the EFT are respectively found to be 
\begin{eqnarray}
	{\rm i} \mathcal{M}^{B,\rm EFT} &=& \frac{{\rm i}\delta^{}_{ab}}{M^2_\Delta} \overline{u} \left( p^{}_2 \right)  P^{}_{\rm R} \left[ \gamma^\mu \slashed{q} \slashed{p}^{}_2 \left(C^\prime_{\widetilde{B}\ell} \right)^{}_{\beta\alpha} + \frac{1}{2} g^{}_1 \gamma^\mu p^2_2 \left( C^{}_{\ell D}\right)^{}_{\beta\alpha} + \gamma^\mu q^2 \left( \frac{1}{4}g^{}_1 C^{}_{\ell D} + C^{}_{B\ell} - \frac{\rm i}{2} C^\prime_{B\ell} \right)^{}_{\beta\alpha} \right.
	\nonumber
	\\
	&& + \gamma^\mu q \cdot p^{}_2 \left( \frac{1}{2} g^{}_1 C^{}_{\ell D} - C^\prime_{\widetilde{B}\ell} - {\rm i} C^\prime_{B\ell} \right)^{}_{\beta\alpha} + g^{}_1 \slashed{p}^{}_2 p^\mu_2 \left( C^{}_{\ell D} \right)^{}_{\beta\alpha} + \slashed{p}^{}_2 q^\mu \left( \frac{1}{2} g^{}_1 C^{}_{\ell D} - C^\prime_{\widetilde{B}\ell} \right)^{}_{\beta\alpha} 
	\nonumber
	\\
	&& + \left. \slashed{q} p^\mu_2 \left( \frac{1}{2} g^{}_1 C^{}_{\ell D} + C^\prime_{\widetilde{B\ell}} + {\rm i} C^\prime_{B\ell} \right)^{}_{\beta\alpha} + \slashed{q} q^\mu \left( \frac{1}{4} g^{}_1 C^{}_{\ell D} - C^{}_{B\ell} + \frac{i}{2} C^\prime_{B\ell} \right)^{}_{\beta\alpha} \right] u \left( p^{}_1 \right) \epsilon^\ast_\mu \left(q\right) \;, \nonumber \\
	\label{eq:B-EFT}
\end{eqnarray}
and 
\begin{eqnarray}
	{\rm i} \mathcal{M}^{W,\rm EFT} &=& \frac{{\rm i}\sigma^J_{ba}}{M^2_\Delta} \overline{u} \left( p^{}_2 \right)  P^{}_{\rm R} \left[ \gamma^\mu \slashed{q} \slashed{p}^{}_2 \left(C^\prime_{\widetilde{W}\ell} \right)^{}_{\beta\alpha} - \frac{1}{2} g^{}_2 \gamma^\mu p^2_2 \left( C^{}_{\ell D}\right)^{}_{\beta\alpha} - \gamma^\mu q^2 \left( \frac{1}{4} g^{}_2 C^{}_{\ell D} - C^{}_{W\ell} + \frac{\rm i}{2} C^\prime_{W\ell} \right)^{}_{\beta\alpha} \right.
	\nonumber
	\\
	&& - \gamma^\mu q \cdot p^{}_2 \left( \frac{1}{2} g^{}_2 C^{}_{\ell D} + C^\prime_{\widetilde{W}\ell} + {\rm i} C^\prime_{W\ell} \right)^{}_{\beta\alpha} - g^{}_2 \slashed{p}^{}_2 p^\mu_2 \left( C^{}_{\ell D} \right)^{}_{\beta\alpha} - \slashed{p}^{}_2 q^\mu \left( \frac{1}{2} g^{}_2 C^{}_{\ell D} + C^\prime_{\widetilde{W}\ell} \right)^{}_{\beta\alpha} 
	\nonumber
	\\
	&& -\left. \slashed{q} p^\mu_2 \left( \frac{1}{2} g^{}_2 C^{}_{\ell D} - C^\prime_{\widetilde{W\ell}} - {\rm i} C^\prime_{W\ell} \right)^{}_{\beta\alpha} - \slashed{q} q^\mu \left( \frac{1}{4} g^{}_2 C^{}_{\ell D} + C^{}_{W\ell} - \frac{\rmI}{2} C^\prime_{W\ell} \right)^{}_{\beta\alpha} \right] u \left( p^{}_1 \right) \epsilon^\ast_\mu \left(q\right) \;.\nonumber \\
		\label{eq:W-EFT}
\end{eqnarray}

Now we proceed to calculate the amplitudes $\langle \ell \ell B \rangle$ and $\langle \ell \ell W \rangle$ in the UV model. It is straightforward to find the corresponding amplitudes for diagrams (a) and (b) in Fig.~\ref{fig:diagrams-matching} for $B^{}_\mu$ as below
\begin{eqnarray}
	{\rm i} \mathcal{M}^{B,\rm UV}_{a} &=& \overline{u} \left(p^{}_2 \right) P^{}_{\rm R} \left\{ \int \frac{{\rm d}^4 k}{\left(2\pi\right)^4} \frac{\left( \slashed{k} - \slashed{p}^{}_2 \right) \gamma^\mu \left(\slashed{k} - \slashed{p}^{}_2 - \slashed{q} \right)}{\left( k^2 - M^2_\Delta \right) \left( k - p^{}_2 \right)^2 \left(k-p^{}_2 - q \right)^2} \right\} u\left( p^{}_1 \right) \epsilon^\ast_\mu \left( q \right)
	\nonumber
	\\
	&& \times \left(- \frac{3}{2} \right) g^{}_1\delta^{}_{ab} \left( Y^{}_\Delta Y^\dagger_\Delta \right)^{}_{\beta\alpha} \;,
	\nonumber
	\\
	{\rm i} \mathcal{M}^{B,\rm UV}_{b} &=& \overline{u} \left(p^{}_2 \right) P^{}_{\rm R} \left\{ \int \frac{{\rm d}^4 k}{\left(2\pi\right)^4} \frac{\left(\slashed{k} - \slashed{p}^{}_2 - \slashed{q} \right) \left( 2k - q \right)^\mu }{\left( k^2 - M^2_\Delta \right)  \left(k-p^{}_2 - q \right)^2 \left[ \left(k - q\right)^2 - M^2_\Delta \right] }  \right\} u\left( p^{}_1 \right) \epsilon^\ast_\mu \left( q \right)
	\nonumber
	\\
	&& \times \left(- 3 \right) g^{}_1\delta^{}_{ab} \left( Y^{}_\Delta Y^\dagger_\Delta \right)^{}_{\beta\alpha} \;.
	\label{eq:B-amp-UV}
\end{eqnarray} 
The hard-momentum parts of these two amplitudes can be obtained by expanding the integrands in the limit of $p \ll k, M^{}_\Delta$ with $p$ being any external momentum (i.e., $p = p^{}_1, p^{}_2, q$ in the present case). Since the combination of the external fields $\ell\ell B$ is already of mass-dimension four, the relevant terms for dim-6 operators should be proportional to the second power of the external momentum. The latter is equivalent to the square of the spacetime derivative. With the help of Eq.~\eqref{eq:B-amp-UV}, we can get the contributions from the hard-momentum region
\begin{eqnarray}
	\left. {\rm i} \mathcal{M}^{B,\rm UV}_{\rm tot} \right|^{}_{\rm hard} &=& \left. {\rm i} \mathcal{M}^{B,\rm UV}_{a} \right|^{}_{\rm hard} + \left. {\rm i} \mathcal{M}^{B,\rm UV}_{b} \right|^{}_{\rm hard} 
	\nonumber
	\\
	&=& \frac{-{\rm i} g^{}_1 \delta^{}_{ab}}{12 \left(4\pi\right)^2 M^2_\Delta} \left( Y^{}_\Delta Y^\dagger_\Delta \right)^{}_{\beta\alpha} \overline{u} \left(p^{}_2 \right) P^{}_{\rm R} \left\{ \gamma^\mu\left[ - 9 \slashed{q} \slashed{p}^{}_2 + 3 p^2_2 +  q^2 \left( 7 + 6 L^{}_\Delta \right) + 12 q \cdot p^{}_2 \right] \right.
	\nonumber
	\\
	&& +\left.  6\left( \slashed{p}^{}_2 - \slashed{q} \right) p^\mu_2 + 2\left[ 6\slashed{p}^{}_2 - \left( 2 + 3L^{}_\Delta \right) q^{}_\mu \right] \right\} u \left( p^{}_1 \right) \epsilon^\ast_\mu \left(q\right) \;,
	\label{eq:B-amp-hard-UV}
\end{eqnarray}
where only the terms proportional to $p^2$ (or equivalently $M^{-2}_\Delta$) have been retained. Similarly, we can obtain the contributions from the hard-momentum region to the amplitude $\langle \ell\ell W \rangle$ in the UV model, that is
\begin{eqnarray}
	\left. {\rm i} \mathcal{M}^{W,\rm UV}_{\rm tot} \right|^{}_{\rm hard} &=& \frac{{\rm i} g^{}_2 \sigma^J_{ba}}{12 \left(4\pi\right)^2 M^2_\Delta} \left( Y^{}_\Delta Y^\dagger_\Delta \right)^{}_{\beta\alpha} \overline{u} \left(p^{}_2 \right) P^{}_{\rm R} \left\{ \gamma^\mu\left[ - 3 \slashed{q} \slashed{p}^{}_2 + 3 p^2_2 +  q^2 \left( 4 + 2 L^{}_\Delta \right) + 6 q \cdot p^{}_2 \right] \right.
	\nonumber
	\\
	&& +\left.  6\slashed{p}^{}_2 p^\mu_2 + \left[ 6\slashed{p}^{}_2 - \left( 1 + 2L^{}_\Delta \right) \slashed{q} \right] q^{}_\mu \right\} u \left( p^{}_1 \right) \epsilon^\ast_\mu \left(q\right)\;.
	\label{eq:W-amp-hard-UV}
\end{eqnarray}
Equating Eq.~\eqref{eq:B-EFT} with Eq.~\eqref{eq:B-amp-hard-UV}, and  Eq.~\eqref{eq:W-EFT} with Eq.~\eqref{eq:W-amp-hard-UV} as well, one can arrive at
\begin{eqnarray}
	&& C^{}_{\ell D} =  - \frac{1}{2\left(4\pi\right)^2} \left(Y^{}_\Delta Y^\dagger_\Delta \right) \;,\; C^\prime_{B\ell} = C^\prime_{W\ell} = 0 \;,\;
	\nonumber
	\\
	&& C^\prime_{\widetilde{B}\ell} = \frac{3g^{}_1}{4\left(4\pi\right)^2} \left(Y^{}_\Delta Y^\dagger_\Delta \right) \;,\; C^\prime_{\widetilde{W}\ell} = -\frac{g^{}_2}{4\left(4\pi\right)^2} \left(Y^{}_\Delta Y^\dagger_\Delta \right) \;,
	\nonumber
	\\
	&& C^{}_{B\ell} = - \frac{g^{}_1}{24\left(4\pi\right)^2} \left( 11 + 12 L^{}_\Delta \right) \left(Y^{}_\Delta Y^\dagger_\Delta \right) \;,\; C^{}_{W\ell} = \frac{g^{}_2}{24\left(4\pi\right)^2} \left( 5 + 4 L^{}_\Delta \right) \left(Y^{}_\Delta Y^\dagger_\Delta \right) \;.
	\label{eq:wc-GB}
\end{eqnarray}
Given the results in Eq.~\eqref{eq:wc-GB} and the connection between the operators in the Green's basis and those in the Warsaw basis, we can obtain the Wilson coefficients of $O^{}_{eB}$ and $O^{}_{eW}$, namely,
\begin{eqnarray}
	C^{}_{eB} &=& - \frac{g^{}_1}{8} C^{}_{\ell D}Y^{}_l - \frac{\rm i}{4} C^\prime_{B\ell} Y^{}_l + \frac{1}{4} C^\prime_{\widetilde{B}\ell} Y^{}_l = + \frac{g^{}_1}{4\left(4\pi\right)^2} \left(Y^{}_\Delta Y^\dagger_\Delta Y^{}_l \right) \;,
	\nonumber
	\\
	C^{}_{eW} &=& +\frac{g^{}_2}{8} C^{}_{\ell D}Y^{}_l - \frac{\rm i}{4} C^\prime_{W\ell} Y^{}_l + \frac{1}{4} C^\prime_{\widetilde{W}\ell} Y^{}_l = - \frac{g^{}_2}{8\left(4\pi\right)^2} \left(Y^{}_\Delta Y^\dagger_\Delta Y^{}_l\right) \;,
	\label{eq:wc-WB}
\end{eqnarray}
which are exactly same as those obtained by the functional approach [cf. Eqs.~(\ref{eq:eBw}) and (\ref{eq:eWw})].

The previous calculations clearly exemplify the diagrammatic approach to one-loop matching. However, owing to the complexity of interactions in the full type-II seesaw model, numerous Feynman diagrams have to be calculated for a complete one-loop matching, so the time-consuming manual calculations seem to be impractical. In our work, we actually implement the diagrammatic approach in a semi-automatic way. First, we use \textsf{FeynRules}~\cite{Christensen:2008} to generate the \textsf{FeynArts}~\cite{Hahn:2001} model file. The interface \textsf{FeynHelper}~\cite{Vladyslav:2016} is utilized to realize the connection among \textsf{FeynArts}, \textsf{FeynCalc}~\cite{Shtabovenk:2016} and \textsf{Package-X}~\cite{Patel:2015}. More explicitly, \textsf{FeynArts} draws all the Feynman diagrams and generates the corresponding amplitudes, \textsf{FeynCalc} calculates the loop integrals involved in the amplitudes, and \textsf{Package-X} automatically converts the amplitudes into Passarino–Veltman functions~\cite{Passarino:1978jh}, whose analytical expressions are provided. All these tools have been implemented in {\sf Mathematica}, and thus it is convenient for us to compare the results from diagrammatic calculations with those from functional approach by {\sf SuperTracer}. Through the Feynman diagrammatic approach, we have cross-checked all the operators up to dim-6 and the Wilson coefficients from {\sf SuperTracer}, and found a complete agreement.

\section{Radiative Decays of Charged Leptons}\label{sec:muegamma}

As a demonstrative example, we shall calculate the branching ratios of radiative decays of charged leptons $l^-_\alpha \to l^-_\beta + \gamma$ in the SEFT-II and compare the results with those calculated in the full type-II seesaw model. In the EFT, the relevant Lagrangian after the spontaneous gauge symmetry breaking is given by 
\begin{eqnarray}
    \mathcal{L}^{}_{\rm SEFT-II} &\supset&  - \overline{l^{}_{\alpha{\rm L}}} \left(M^{}_l\right)^{}_{\alpha\beta} l^{}_{\beta{\rm R}} - \frac{1}{2} \overline{\nu^{}_{\alpha{\rm L}}} \left(M^{}_\nu\right)^{}_{\alpha\beta} \nu^{\rm c}_{\beta{\rm L}} + \frac{g^{}_2}{\sqrt{2}} \overline{l^{}_{\alpha{\rm L}}} \gamma^\mu \nu^{}_{\alpha{\rm L}} W^-_\mu %+ {\rm h.c.} 
    \nonumber
    \\
    && + \frac{v}{\sqrt{2}M^2_\Delta} \left( \cos\theta^{}_{\rm w} C^{\alpha\beta}_{eB} - \sin\theta^{}_{\rm w} C^{\alpha\beta}_{eW} \right) \overline{l^{}_{\alpha \rm L}} \sigma^{}_{\mu\nu} l^{}_{\beta\rm R} F^{\mu\nu} + {\rm h.c.} \; ,
    \label{eq:lagrangian-ssb}
\end{eqnarray}
where only the terms related to radiative decays of charged leptons are kept. In Eq.~(\ref{eq:lagrangian-ssb}), we have defined the charged-lepton mass matrix $M^{}_l \equiv v Y^{}_l/\sqrt{2}$ and the effective neutrino mass matrix $M^{}_\nu \equiv v^2 \lambda^{}_\Delta Y^{}_\Delta /M^{}_\Delta$ induced by the dim-5 Weinberg operator. The electromagnetic dipole operator in the second line of Eq.~\eqref{eq:lagrangian-ssb} results from the two dim-6 operators $O^{}_{eB}$ and $O^{}_{eW}$, for which the Wilson coefficients are respectively $C^{}_{eB} = g^{}_1 \left( Y^{}_\Delta Y^\dagger_\Delta Y^{}_l \right)/(64\pi^2)$ and $C^{}_{eW} = -g^{}_2 \left( Y^{}_\Delta Y^\dagger_\Delta Y^{}_l \right)/(128\pi^2)$. In addition, $\theta^{}_{\rm w} = \arctan(g^{}_1/g^{}_2)$ is the weak mixing angle, and $F^{}_{\mu\nu} = \partial^{}_\mu A^{}_\nu - \partial^{}_\nu A^{}_\mu$ is the gauge field strength with $A^{}_\mu$ being the photon field. Note that the terms in the first line of Eq.~\eqref{eq:lagrangian-ssb} appear at the tree level while those in the second line at the one-loop level.

Working in the flavor basis where the charged-lepton mass matrix is diagonal, namely, $M^{}_l = {\rm Diag} \{ m^{}_e, m^{}_\mu, m^{}_\tau \}$, we can make a field transformation $\nu^{}_{\rm L} \to U \nu^{}_{\rm L}$ with $U$ being the unitary matrix to diagonalize the effective neutrino mass matrix via $U^\dagger M^{}_\nu U^\ast = \widehat{M}^{}_\nu = {\rm Diag} \{ m^{}_1, m^{}_2, m^{}_3 \}$. Then the Lagrangian in Eq.~\eqref{eq:lagrangian-ssb} turns out to be
%%%%%%%%%%%%%%%%%%%%%%%%%%figure 3 %%%%%%%%%%%%%%%%%%%%%%%%%%
\begin{figure}[t]
\centering
\includegraphics[width=1\textwidth]{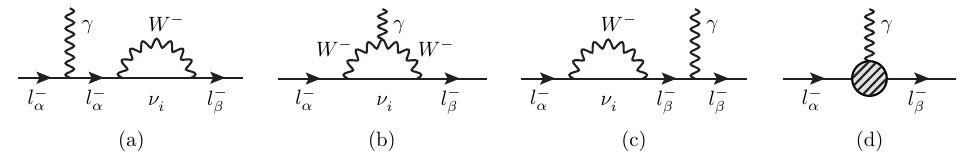}
\vspace{-0.3cm}
\caption{Feynman diagrams for radiative decays of charged leptons $l^-_\alpha \to l^-_\beta + \gamma$ at one-loop level in the EFT. The
unitary gauge has been adopted. (a)-(c) are mediated by massive neutrinos via the charged current interactions of leptons, (d) is generated by the dimension-six operators at one-loop level.}
\label{fig:diagrams-eft}
\end{figure}
%%%%%%%%%%%%%%%%%%%%%%%%%%%%%%%%%%%%%%%%%%%%%%%%%%%%%%%%%%%%%
%%%%%%%%%%%%%%%%%%%%%%%%%%figure 4 %%%%%%%%%%%%%%%%%%%%%%%%%%
\begin{figure}[t]
\centering
\includegraphics[width=1\textwidth]{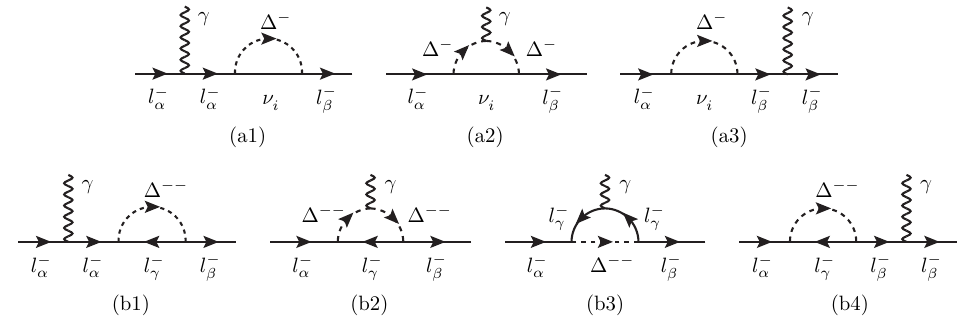}
\vspace{-0.3cm}
\caption{Feynman diagrams mediated by the singly- and doubly-charged scalars for radiative decays of charged leptons $l^-_\alpha \to l^-_\beta + \gamma$ at the one-loop level in the full type-II seesaw model. The unitary gauge has been adopted.}
\label{fig:diagrams-uv}
\end{figure}
%%%%%%%%%%%%%%%%%%%%%%%%%%%%%%%%%%%%%%%%%%%%%%%%%%%%%%%%%%%%%
\begin{eqnarray}
    \mathcal{L}^{}_{\rm SEFT-II} &\supset& - \overline{l^{}_{\rm L}} M^{}_l l^{}_{\rm R} - \frac{1}{2} \overline{\nu^{}_{\rm L}} \widehat{M}^{}_\nu \nu^{\rm c}_{\rm L} + \frac{g^{}_2}{\sqrt{2}} \overline{l^{}_{\rm L}} \gamma^\mu U \nu^{}_{\rm L} W^-_\mu %+ {\rm h.c.} 
    \nonumber
    \\
    && + \frac{3e}{8(4\pi)^2M^2_\Delta} \overline{l^{}_{\rm L}} \sigma^{}_{\mu\nu} Y^{}_\Delta Y^\dagger_\Delta M^{}_l l^{}_{\rm R} F^{\mu\nu} + {\rm h.c.} \; ,
    \label{eq:lagrangian-diag}
\end{eqnarray}
in which $e=g^{}_1 \cos\theta^{}_{\rm w} = g^{}_2 \sin\theta^{}_{\rm w}$ has been used, and the unitary matrix $U$ is the Pontecorvo-Maki-Nakagawa-Sakata (PMNS) mixing matrix~\cite{Pontecorvo:1957cp,Maki:1962mu,Pontecorvo:1967fh}. From the Lagrangian given in Eq.~\eqref{eq:lagrangian-diag}, one can observe that there are two kinds of contributions to the branching ratios of $l^-_\alpha \to l^-_\beta + \gamma$ in the EFT. One is from massive neutrinos via one-loop diagrams induced by the charged-current interactions of leptons, i.e., the diagrams (a)-(c) in Fig.~\ref{fig:diagrams-eft}. The other one comes directly from the dim-6 operators $O^{}_{eB}$ and $O^{}_{eW}$ [i.e., the electromagnetic dipole operator in the second line of Eq.~\eqref{eq:lagrangian-diag}], corresponding to the diagram (d) in Fig.~\ref{fig:diagrams-eft}. In fact, the former contribution is entirely included in the full theory, whereas the latter corresponds to the contributions from the singly- and doubly-charged scalars $\Delta^-$ and $\Delta^{--}$ in the full theory. The triplet-mediated diagrams have been shown in Fig.~\ref{fig:diagrams-uv}. 

The radiative decays of charged leptons induced by massive neutrinos [i.e., the processes corresponding to diagrams (a)-(c) in Fig.~\ref{fig:diagrams-eft}] have been investigated a long time ago~\cite{Minkowski:1977sc,Petcov:1976ff,Bilenky:1977du,Cheng:1976uq,Marciano:1977wx,Lee:1977qz,Lee:1977tib}. The amplitudes of these rare decays are highly suppressed due to the smallness of neutrino masses, and to the unitarity of the PMNS matrix (i.e., the Glashow-Iliopoulos-Maiani mechanism~\cite{Glashow:1970gm} in the leptonic sector) as well. As a consequence, in the type-II seesaw model, the contributions given by the diagrams (a)-(c) in Fig.~\ref{fig:diagrams-eft} are usually omitted and only those from $\Delta^-$ and $\Delta^{--}$ shown in Fig.~\ref{fig:diagrams-uv} need to be taken into account.\footnote{If the masses of triplet scalars are very large, the contributions from massive neutrinos via diagrams (a)-(c) in Fig.~\ref{fig:diagrams-eft} may be comparable to those from $\Delta^-$ and $\Delta^{--}$ via the diagrams in Fig.~\ref{fig:diagrams-uv}. In this case, all the contributions should be considered though all of them are very small.} Similarly,  we shall also ignore the contributions from massive neutrinos to the radiative decays of charged leptons, and focus only on those from the electromagnetic dipole operator in the second line of Eq.~\eqref{eq:lagrangian-diag}. It is easy to obtain the amplitude corresponding to the diagram (d) in Fig.~\ref{fig:diagrams-eft}:
\begin{eqnarray}
	{\rm i} \mathcal{M}  = \frac{3e}{4\left(4\pi\right)^2 M^2_\Delta} \left(Y^{}_\Delta Y^\dagger_\Delta \right)^{}_{\beta\alpha} \overline{u} \left( p^{}_2 \right)  \sigma^{\mu\nu}q^{}_\nu \left( m^{}_\alpha P^{}_{\rm R} + m^{}_\beta P^{}_{\rm L} \right) u\left(p^{}_1\right) \epsilon^\ast_\mu \left(q\right) \;.\qquad
	\label{eq:amp}
\end{eqnarray}
Then, the decay rate is given by 
\begin{eqnarray}
	\Gamma \left( l^-_\alpha \to l^-_\beta + \gamma \right) &=& \frac{1}{2m^{}_\alpha} \cdot \frac{1}{8\pi} \left( 1- \frac{m^2_\beta}{m^2_\alpha} \right) \cdot \frac{1}{2} \sum \left| \mathcal{M} \right|^2
	\nonumber
	\\
	&=& \frac{9\alpha^{}_{\rm em}m^5_\alpha}{64 \left( 4\pi\right)^4 M^4_\Delta} \left( 1 + \frac{m^2_\beta}{m^2_\alpha} \right) \left( 1 - \frac{m^2_\beta}{m^2_\alpha} \right)^3 \left|\left( Y^{}_\Delta Y^\dagger_\Delta \right)^{}_{\beta\alpha} \right|^2 
	\nonumber
	\\
	&\simeq&  \frac{9\alpha^{}_{\rm em}m^5_\alpha}{64 \left( 4\pi\right)^4 M^4_\Delta} \left|\left( Y^{}_\Delta Y^\dagger_\Delta \right)^{}_{\beta\alpha} \right|^2 \;,
	\label{eq:decay-ratio}
\end{eqnarray}
where $\alpha^{}_{\rm em} \equiv e^2/(4\pi)$ is the electromagnetic fine-structure constant, and the tiny ratio $m^2_\beta/m^2_\alpha \ll 1$ in the last step has been neglected. For convenience, one can define a dimensionless ratio between the rate of radiative decays and that of purely leptonic decays $l^-_\alpha \to l^-_\beta + \overline{\nu^{}_\beta} + \nu^{}_{\alpha}$, that is 
\begin{eqnarray}
	\xi\left( l^-_\alpha \to l^-_\beta + \gamma \right) \equiv \frac{\Gamma \left( l^-_\alpha \to l^-_\beta + \gamma \right)}{\Gamma \left( l^-_\alpha \to l^-_\beta + \overline{\nu^{}_\beta} + \nu^{}_{\alpha} \right)} \simeq \frac{27\alpha^{}_{\rm em}}{ 256\pi G^2_{\rm F} M^4_\Delta} \left|\left( Y^{}_\Delta Y^\dagger_\Delta \right)^{}_{\beta\alpha} \right|^2 \;,
	\label{eq:ratio-eft}
\end{eqnarray}
where $G^{}_{\rm F}$ is the Fermi constant. For $\mu \to e \gamma$, the dimensionless ratio is approximately the branching ratio, i.e., ${\rm BR} \left( \mu \to e \gamma \right) \simeq \xi \left( \mu \to e \gamma \right)$, as the purely leptonic decay $\mu^- \to e^- + \overline{\nu^{}_\mu} + \nu^{}_\mu$ dominates over all other channels. One can make use of Eq.~\eqref{eq:ratio-eft} to constrain the Yukawa coupling matrix $Y^{}_\Delta$ if the experimental constraints on the radiative decays of charged leptons are taken into account.

The radiative decay of muon $\mu \to e \gamma$ in the type-II seesaw model has also been extensively calculated~\cite{Cuypers:1996ia, Kakizaki:2003jk, Akeroyd:2009nu, Fukuyama:2009xk, Chakrabortty:2012vp, Dinh:2012bp, Lindner:2016bgg}. Once the diagrams in Fig.~\ref{fig:diagrams-uv} are evaluated, the branch ratio of $\mu \to e \gamma$ contributed from $\Delta^-$ and $\Delta^{--}$ will be~\cite{Cuypers:1996ia, Kakizaki:2003jk, Akeroyd:2009nu, Fukuyama:2009xk, Chakrabortty:2012vp, Dinh:2012bp, Lindner:2016bgg}\footnote{Note that different conventions of the Yukawa coupling matrix exist in the literature, namely, $Y^{}_\Delta$ in our work and $h$ in some references. The connection between these two conventions is given by $h=Y^\dagger_\Delta/\sqrt{2}$.  } 
\begin{eqnarray}
	{\rm BR}^\prime \left( \mu \to e \gamma \right) &\simeq& \frac{\alpha^{}_{\rm em}}{192 \pi G^2_{\rm F}}  \left( \frac{1}{M^2_{\Delta^{-}}} + \frac{8}{M^2_{\Delta^{--}}} \right)^2 \cdot \frac{1}{4}\left| \left( Y^{}_\Delta Y^\dagger_\Delta \right)^{}_{e\mu} \right|^2 
	\nonumber
	\\
	&\simeq& \frac{27\alpha^{}_{\rm em}}{ 256\pi G^2_{\rm F} M^4_\Delta} \left|\left( Y^{}_\Delta Y^\dagger_\Delta \right)^{}_{e\mu} \right|^2 \;.
	\label{eq:ratio-full}
\end{eqnarray}
In the second line of Eq.~\eqref{eq:ratio-full}, the mass spectrum $M^2_{\Delta^{--}} = M^2_\Delta$ and $M^2_{\Delta^{-}}= M^2_\Delta \left( 1 + 2\lambda^2_\Delta v^2/M^2_\Delta \right) \simeq M^2_\Delta$ has been considered. 

As is expected, the results in Eqs.~\eqref{eq:ratio-eft} and \eqref{eq:ratio-full} are completely consistent with each other for the radiative decay of muon $\mu \to e \gamma$. In a similar way, one can
check other decay channels.

\section{Summary}\label{sec:summary}

In the present paper, we have accomplished a complete one-loop matching of the type-II seesaw model onto the SMEFT. The primary motivation for the construction of the low-energy EFT for the type-II seesaw model is two-fold. First, neutrino oscillations have provided us with very strong evidence that neutrinos are massive particles. The origin of neutrino masses definitely calls for a renormalizable UV theory beyond the SM. As one of the simplest and most natural models for tiny Majorana neutrino masses, the type-II seesaw model has to be scrutinized by experimental tests. Second, as the particle physics has entered the precision era, more accurate data require more precise calculations. The latter could be either the loop-level matching condition between the UV theory and the EFT or the higher-order calculations in the EFT. The one-loop matching of the type-II seesaw model onto the SMEFT sets up a self-consistent framework to investigate low-energy phenomenology of the EFT and to analyze experimental results of possible deviations from the SM predictions.

The functional approach has been implemented in this work to derive all the effective operators up to dim-6 in the Warsaw basis and the corresponding Wilson coefficients. In addition, the one-loop threshold corrections to the mass parameter and the couplings in the SM and also to the coefficient of the dim-5 operator are given up to $\mathcal{O}\left( M^{-2} \right)$, which are the very matching conditions for the two-loop RGEs of these physical parameters. At the one-loop level, we have found 41 dim-6 operators in the SEFT-II, covering all the 31 dim-6 operators in the SEFT-I. Ten dim-6 operators present in the SEFT-II but not in the SEFT-I arise solely from the gauge interactions of the Higgs triplet in the type-II seesaw model. It is worthwhile to mention that the dim-6 operators and their Wilson coefficients derived by the functional approach have been cross-checked by the diagrammatic approach. The calculations by using both functional and diagrammatic approaches are necessary to ensure the correctness of the final results.

Finally, we point out that there are several important issues left for further studies. First, the EFT has been constructed by one-loop matching at the decoupling scale characterized by the masses of heavy degrees of freedom. The two-loop RGEs of the SM parameters and the Wilson coefficients in the EFT must be derived and used to run all the physical parameters from the high-energy scale to the low-energy one. Second, now that the EFTs for both type-I and type-II seesaw models are available, it is interesting to explore the distinct experimental signatures induced by the dim-6 operators that are only present in the SEFT-II. Third, once the EFTs at the low-energy scale are obtained, a global-fit analysis of all relevant experimental data in the SEFT-II framework will be indispensable to probe the fundamental parameters in the UV theory. We hope to come back to these issues in the near future.

\section*{Acknowledgements}
The authors thank Prof. Zhi-zhong Xing for helpful discussions. This work was supported in part by the National Natural Science Foundation of China under grant No.~11835013 and No.~12075254, by the Key Research Program of the Chinese Academy of Sciences under grant No. XDPB15, and by the CAS Center for Excellence in Particle Physics.

\section*{Note added}

After completing one-loop matching of the type-II seesaw model onto the SMEFT, we are kindly informed that a similar work has also been done independently in Ref.~\cite{Yu}. When our paper is being finalized, the package {\sf Matchmakereft} for one-loop matching via diagrammatic approach is released~\cite{Carmona:2021xtq}. In this updated version, we have cross-checked our results by using {\sf Matchmakereft} and found an excellent agreement except for the additional contribution in Eq.~(\ref{eq:second-derivative}) to the Wilson coefficient for the operator $O^{}_H$ [cf. Eq.~(\ref{eq:H6})]. Such a discrepancy clearly originates from the simple implementation of EOMs in changing from the Green's basis to the Warsaw basis in Ref.~\cite{Carmona:2021xtq}, as we have explained in Sec.~\ref{sec:4.1}.

\end{document}